\documentclass[aps,pra,twocolumn,10pt,superscriptaddress,longbibliography,nobibnotes,nofootinbib,floatfix]{revtex4-2}
\usepackage{amsmath,amssymb,graphicx,bm}
\usepackage{tikz}
\usetikzlibrary{arrows.meta,decorations.pathmorphing,positioning,calc}
\usepackage[colorlinks=true,allcolors=blue]{hyperref}
\usepackage{orcidlink}

\newcommand{\Sp}{\hat S^+}
\newcommand{\Sm}{\hat S^-}

\newcommand{\Sy}{\hat S^y}
\newcommand{\Sz}{\hat S^z}
\newcommand{\ah}{\hat a}
\newcommand{\ad}{\hat a^\dagger}
\newcommand{\nm}{\hat n^{m}}
\newcommand{\nb}{\hat n}
\newcommand{\Hh}{\hat H}
\newcommand{\rhoh}{\hat\rho}
\newcommand{\Ident}{\hat{\mathbb{I}}}
\newcommand{\ket}[1]{|#1\rangle}
\newcommand{\bra}[1]{\langle#1|}
\newcommand{\braket}[2]{\langle#1|#2\rangle}
\newcommand{\dyad}[1]{|#1\rangle\langle#1|}
\newcommand{\FM}{\Uparrow}
\DeclareMathOperator{\Tr}{Tr}
\newcommand{\FQ}{\mathcal{F}_Q}
\newcommand{\Fglob}{\mathcal{F}_{\rm global}}
\newcommand{\Fs}{\mathcal{F}^{\rm s}}
\newcommand{\Fb}{\mathcal{F}^{\rm b}}
\newcommand{\Fsb}{\mathcal{F}^{\rm sb}}
\newcommand{\dFsb}{\Delta\mathcal{F}^{\rm sb}}
\newcommand{\Fsbar}{\bar{\mathcal{F}}^{\rm s}}
\newcommand{\Fbbar}{\bar{\mathcal{F}}^{\rm b}}
\newcommand{\dFsbbar}{\Delta\bar{\mathcal{F}}^{\rm sb}}
\newcommand{\Nbtot}{\langle N^{\rm b}\rangle}
\newcommand{\Nbcond}{\bar n^{\rm b}}

\begin{document}

\title{Flow of local sensitivity in a spin chain coupled to a bosonic bath}

\author{Marcin P\l odzie\'n\,\orcidlink{0000-0002-0835-1644}}
\affiliation{Qilimanjaro Quantum Tech, Carrer de Vene\c{c}uela 74, 08019 Barcelona, Spain}

\begin{abstract}
We study how local sensitivity to an encoded parameter, captured by the quantum Fisher information, flows between a spin chain, a coupled bosonic bath, and their quantum correlations, where we treat the bath as a full many-body quantum system beyond the Lindbladian approximation. We analyze how the coupling symmetry determines the destination of the departed sensitivity directly at the level of the Hamiltonian. We consider an excitation-exchanging Jaynes--Cummings coupling, which preserves the total number of excitations, and a spin-excitation-conserving Holstein coupling. We find that the Holstein coupling leaves the bath with no first-order information about the phase and stores the lost sensitivity entirely in spin--bath correlations, whereas the Jaynes--Cummings coupling passes this sensitivity to the bath, in full for a single excitation. The bath spectrum then governs whether the sensitivity ever returns to the spins. It revives fully and periodically when the coupled-system frequencies share a common period, but when those frequencies disperse the departed share dephases across the bath modes and never returns. We note that a transported metrological register loses more sensitivity than its arrival fidelity implies.
\end{abstract}

\maketitle


\section{Introduction}\label{sec:intro}
The quantum Fisher information sets the fundamental precision limit for parameter estimation~\cite{Braunstein1994,Paris2009,Pezze2018,Toth2014} and the maximum gain available from entangled probes~\cite{Giovannetti2004,Giovannetti2011}. Coupling to an environment degrades this precision. This metrological cost is typically evaluated through precision bounds and noise channels~\cite{Escher2011,Demkowicz2012,Smirne2016}, or followed dynamically using a master equation for the probe alone~\cite{BreuerPetruccione2002}. In the latter Markovian approach, the environment often enters merely as a set of decay rates. Both methods successfully describe how fast a probe degrades. Neither approach can capture where the lost sensitivity actually went because the environment is traced out.

In this work we follow the complete time evolution of local sensitivity in a spin chain coupled to a bosonic bath. We treat the bath as an explicit quantum many-body system and evolve it together with the probe instead of reducing the dynamics to a master equation. The joint evolution is unitary, making the loss of information entirely coherent. Any sensitivity that leaves the spins must therefore reside in the bath or in the spin--bath correlations. We trace where this information sits at each moment and identify which measurements can still recover it. The return of this sensitivity is a memory effect of the bath~\cite{deVega2017,Breuer2009,Rivas2010,Lu2010,Chin2012,Rivas2014,Breuer2016}. It revives fully and periodically when the coupled-system frequencies share a common period, but otherwise the departed share dephases across the modes and never returns. By mapping these dynamics we map the conditions for a full revival to the spins. Since the Hamiltonian does not depend on the encoded parameter, the unitary joint evolution conserves the total quantum Fisher information, so that any loss from the spins matches the information absorbed by the bath or the correlations. We initialize the bath in its vacuum and maintain the joint evolution at absolute zero temperature, which fixes the conserved total at the pure-state value set by the generator alone.

Our setting is a spin chain that carries a locally encoded phase while coupled to a tunable bosonic bath. Experimental capabilities restrict which regions can be measured, making the Fisher information of a reduced state defined on a specific spatial block the natural quantity to follow. The dynamics of local sensitivity encoded in isolated spin chains has been studied in Refs.~\cite{Wysocki2024,Wysocki2025,Plodzien2026c}, and block measures of the quantum Fisher information, read out from individual spins or localized regions of the chain, were developed in Refs.~\cite{Plodzien2025,Wysocki2026,Plodzien2026a,Plodzien2026b}. A phase transported across such a chain reaches distant spins along a light cone bounded by the Lieb--Robinson velocity~\cite{Wysocki2024}, and it stays entirely among the spins even when a continuous symmetry of the chain is broken~\cite{Plodzien2026c}. Introducing the bath adds an external party to this dynamic, acting as a reservoir that can either hold the transported sensitivity or return it.

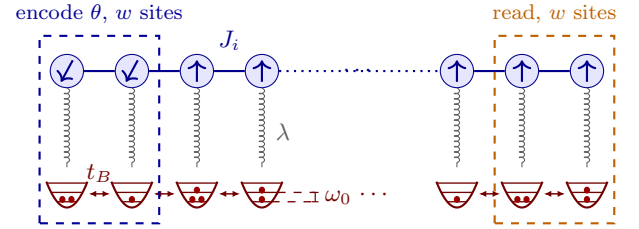
\begin{figure}[t!]
  \centering
  \begin{tikzpicture}[
    x=0.86cm, y=0.95cm, font=\small,
    spin/.style={circle, draw=blue!55!black, fill=blue!10, minimum size=4.4mm, inner sep=0pt},
    exch/.style={thick, blue!55!black},
    hop/.style={thick, red!45!black},
    lev/.style={red!45!black, line width=0.5pt},
    cpl/.style={gray!75!black, decorate, decoration={coil, amplitude=1.0pt, segment length=2.0pt}},
  ]
    \def\ywell{-0.95}\def\awell{3.6}\def\wtop{0.36}
    \foreach \i in {0,1,2,3,6,7,8}{
      \draw[hop] plot[domain=-0.316:0.316, samples=31, smooth] (\i+\x, {\ywell + \awell*\x*\x});
      \foreach \h in {0.08,0.21,0.34}{
        \pgfmathsetmacro\hw{sqrt(\h/\awell)}
        \draw[lev] (\i-\hw, \ywell+\h) -- (\i+\hw, \ywell+\h);
      }
    }
    \newcommand{\bos}[2]{\fill[red!60!black] (#1,\ywell+#2+0.032) circle (0.048);}
    \bos{-0.07}{0.08}\bos{0.07}{0.08}                 
    \bos{1.0}{0.08}                                    
    \bos{1.93}{0.08}\bos{2.07}{0.08}\bos{2.0}{0.21}    
    \bos{3.0}{0.08}\bos{3.0}{0.21}                     
    \bos{6.0}{0.08}                                    
    \bos{6.93}{0.08}\bos{7.07}{0.08}                   
    \bos{8.0}{0.08}\bos{8.0}{0.21}                     
    \draw[lev, dashed] (3.149,\ywell+0.08) -- (3.82,\ywell+0.08);
    \draw[lev, dashed] (3.242,\ywell+0.21) -- (3.82,\ywell+0.21);
    \draw[red!45!black, line width=0.45pt] (3.82,\ywell+0.08) -- (3.82,\ywell+0.21);
    \draw[red!45!black, line width=0.45pt] (3.77,\ywell+0.08) -- (3.87,\ywell+0.08);
    \draw[red!45!black, line width=0.45pt] (3.77,\ywell+0.21) -- (3.87,\ywell+0.21);
    \node[red!45!black, anchor=west, inner sep=1.8pt] at (3.89,\ywell+0.145) {$\omega_0$};
    \foreach \i/\j in {0/1,1/2,2/3,6/7,7/8}{
      \draw[hop, line width=0.55pt,
            {Latex[length=1.0mm,width=0.85mm]}-{Latex[length=1.0mm,width=0.85mm]}]
            (\i+0.335,\ywell+0.19) -- (\j-0.335,\ywell+0.19);
    }
    \node[red!45!black] at (4.75,\ywell+0.18) {$\cdots$};
    \foreach \i in {0,1,2,3,6,7,8}{
      \node[spin] (s\i) at (\i,0.95) {};
      \draw[cpl] (s\i) -- (\i,\ywell+0.56);
    }
    \foreach \i/\j in {0/1,1/2,2/3,6/7,7/8}{
      \draw[exch] (s\i) -- (s\j);
    }
    \draw[exch, dotted] (s3) -- (s6);
    \node[blue!55!black] at (4.5,0.95) {$\cdots$};
    \foreach \i in {2,3,6,7,8}{\draw[->,thick,blue!55!black] ($(s\i)+(0,-0.15)$) -- ++(0,0.30);}
    \foreach \i in {0,1}{\draw[->,thick,blue!55!black] ($(s\i)+(0.09,0.15)$) -- ++(-0.18,-0.30);}
    \node[blue!55!black] at (2.5,1.38) {$J_i$};
    \node[red!45!black, anchor=south] at (0.5,\ywell+0.235) {$t_B$};
    \node[gray!75!black, anchor=west] at (3.08,0.10) {$\lambda$};
    \draw[dashed, thick, blue!60!black] (-0.42,\ywell-0.22) rectangle (1.42,1.44);
    \node[blue!60!black, anchor=south] at (0.5,1.50) {\footnotesize encode $\theta$, $w$ sites};
    \draw[dashed, thick, orange!75!black] (6.58,\ywell-0.22) rectangle (8.42,1.44);
    \node[orange!75!black, anchor=south] at (7.5,1.50) {\footnotesize read, $w$ sites};
  \end{tikzpicture}
  \caption{A schematic illustration of the physical model. A spin chain with uniform or engineered exchange coupling (blue lines) couples to a local bosonic mode per site (red levels, featuring a local frequency and nearest-neighbor hopping). A parameter phase is encoded on a localized block of sites (blue dashed rectangle) and read out at later times from another restricted region of sites (orange dashed rectangle).}
  \label{fig:schematic}
\end{figure}

In this work, we trace how this locally encoded quantum Fisher information partitions across space and time. Two primary factors govern this partition. First, the symmetry of the coupling operator governs the destination of the departed sensitivity. This symmetry alone decides whether measuring the bath can recover the phase at first order or if a joint reading of both parties is required. Second, the bath spectrum controls the return of the sensitivity. For entangled multiparticle probes, the coupling symmetry adds an extra constraint that decides whether a revival can occur at all.

We find that the correlation deficit grows significantly as a bath mode softens toward a vanishing frequency. In this regime the system approaches a limit where the bath traps the sensitivity permanently. We also note that when a metrological register is transported across the chain, its block quantum Fisher information decays faster than its arrival fidelity implies.

The remainder of this paper is structured as follows. In Sec.~\ref{sec:model} we define the physical model and our localized measures of quantum Fisher information. In Sec.~\ref{sec:dynamics} we analyze single-excitation dynamics, and in Sec.~\ref{sec:ghz} we examine collective entangled probes. Next, in Sec.~\ref{sec:pst}, we investigate probe transport on engineered chains, and we provide concluding remarks in Sec.~\ref{sec:concl}. All solutions are obtained via exact diagonalization where applicable; for itinerant and transport problems we use matrix-product-state methods and cross-check them against independent numerical techniques.

\section{Model}\label{sec:model}
We consider $L$ lattice sites, each carrying a spin $\tfrac12$ and a local bosonic mode. The Hamiltonian is
\begin{equation}\label{eq:H}
  \Hh = \Hh_{\rm s} + \Hh_{\rm b} + \lambda\,\Hh_{\rm int},
\end{equation}
with overall coupling strength $\lambda$, a spin XX exchange (excitation hopping),
\begin{equation}
  \Hh_{\rm s} = -\frac{J}{2}\sum_{i=1}^{L-1}\big(\Sp_i\Sm_{i+1} + {\rm h.c.}\big),
\end{equation}
a noninteracting bosonic bath with on-site energy $\omega_0$ and hopping $t_B$,
\begin{equation}
  \Hh_{\rm b} = \omega_0\sum_{i=1}^{L} \nb_i - t_B\sum_{i=1}^{L-1}\big(\ad_i\ah_{i+1}+{\rm h.c.}\big),
\end{equation}
and one of two system--bath coupling operators $\Hh_{\rm int}$,
\begin{subequations}\label{eq:couplings}
\begin{align}
  \Hh_{\rm int}^{\rm JC}  &= \sum_i\big(\Sp_i\ad_i + \Sm_i\ah_i\big),  \label{eq:jc}\\
  \Hh_{\rm int}^{\rm Hol} &= \sum_i \nm_i\,(\ah_i+\ad_i),              \label{eq:hol}
\end{align}
\end{subequations}
where $\nm_i=\tfrac12-\Sz_i\in\{0,1\}$ counts spin excitations, an excitation being a down spin on the fully polarized
background $\ket{\FM}=\bigotimes_i\ket{\uparrow}_i$, with $\ket{j}=\Sm_j\ket{\FM}$. The bath starts in the
vacuum $\ket{\varnothing}=\bigotimes_i\ket{0;i}$, so the spin excitation is the sole
source of excitation. The bath band runs from $\omega_0-2t_B$ to $\omega_0+2t_B$ and stays positive only
for $\omega_0>2t_B$; the Holstein simulations are restricted to that range. The Jaynes--Cummings coupling is not restricted in this way. Because it conserves the total excitation
number, a single excitation spans a finite space whatever the band does, and the resonant illustration
below at $\omega_0=0$ stays well posed even though the band is then partly negative. We set $\hbar=1$ and measure every energy, rate and inverse time in units of the
exchange, $J=1$.

The reservoir is finite and structured, one mode per site with hopping $t_B$. At $t_B=0$ each mode is an
isolated oscillator and the dynamics recurs; decay sets in only for $t_B\neq0$, and at finite $L$ it is
effective rather than strict, since the departed share returns on the boundary-reflection time. Every
window used below sits inside that time.

The spins carry no field term, so their transition frequency is zero and $\omega_0$ is the spin--boson
detuning. The Jaynes--Cummings coupling depends on $\omega_0$ only through the detuning $\delta_q$ introduced in
Sec.~\ref{sec:jc}, which makes the resonant point $\omega_0=0$ used below well defined, with
$\lambda/\omega_0$ acting as the relevant ratio of coupling to detuning away from resonance. The Holstein coupling does not commute with
$\Hh_{\rm b}$ and no rotating frame removes $\omega_0$; there $\omega_0$ is the mode frequency itself and
$g=(\lambda/\omega_0)^2$ measures a genuine polaronic dressing. The Holstein coupling is longitudinal, so
no rotating-wave approximation enters.

The Holstein coupling, from the molecular-crystal
model~\cite{Holstein1959}, conserves the spin-excitation number and reaches the bath only through the on-site density;
the Jaynes--Cummings coupling, from the emitter--mode exchange of quantum optics~\cite{JaynesCummings1963}
(its lattice version here is of the Jaynes--Cummings--Hubbard type),
conserves instead the total excitation number and trades the spin excitation for a boson outright; we
retain its $U(1)$ symmetry, as in setups without counter-rotating terms. Frozen on a single
site, the Holstein model is the exactly solvable independent-boson model of pure
dephasing~\cite{Duke1965,Mahan2000,Palma1996}.

\subsection{Encoding and the QFI}
At $t=0$ a phase $\theta$ is encoded by $\hat U_\theta=e^{-i\theta\hat G}$ with generator
$\hat G=\sum_{j=k}^{k+w-1}\Sy_j$ (the leftmost sites, $k=1$), so that
$\ket{\Psi(\theta,t)}=e^{-i\Hh t}\hat U_\theta\ket{\FM}\ket{\varnothing}$. The encoding introduces
spin excitations into the chain, which carry the phase as they propagate. The QFI of a state $\rhoh(\theta)$
is
\begin{equation}\label{eq:qfi}
  \FQ[\rhoh] = 2\sum_{p_i+p_j>0}\frac{|\bra{i}\partial_\theta\rhoh\ket{j}|^2}{p_i+p_j},
\end{equation}
with $p_i,\ket{i}$ the eigensystem of $\rhoh$. The evolution of the joint chain and bath is unitary and the
global state stays pure, so the QFI of the whole system is set by the generator alone,
$\Fglob=4\,\mathrm{Var}_{\Psi_0}(\hat G)$, the same for every coupling and at every time, so whatever the bath
does redistributes a fixed total [App.~\ref{app:method}]. A phase encoded upstream is read out downstream from the spins alone, from the bath alone, or from a joint
block, and the block QFI is the precision that such a restricted readout can reach. Following it in time shows which restricted measurement still retains the encoded sensitivity.

We read the QFI from finite regions of the chain. For a
contiguous block of $w$ sites $B^{(w)}_j=\{j,\dots,j+w-1\}$ we consider the spin, boson, and joint spin--boson
reduced states
\begin{subequations}\label{eq:rdm_block}
\begin{align}
  \rhoh^{{\rm s},(w)}_j &= \Tr_{\rm bath}\,\Tr_{\overline{B}}\,\dyad{\Psi}, \\
  \rhoh^{{\rm b},(w)}_j &= \Tr_{\rm spin}\,\Tr_{\overline{B}}\,\dyad{\Psi}, \\
  \rhoh^{{\rm sb},(w)}_j &= \Tr_{\overline{B}}\,\dyad{\Psi},
\end{align}
\end{subequations}
which trace out, respectively, the bath and the spins outside the block, all spins together with the modes
outside the block, and only the sites outside the block, with $\overline{B}$ the complement of $B^{(w)}_j$.
The $w=1$ spin state is the single-qubit reduced state at site $j$. Their QFIs are the local densities
\begin{equation}\label{eq:fdensities}
\begin{split}
  \Fs(j,t;w)&=\FQ[\rhoh^{{\rm s},(w)}_j],\\
  \Fb(j,t;w)&=\FQ[\rhoh^{{\rm b},(w)}_j],\\
  \Fsb(j,t;w)&=\FQ[\rhoh^{{\rm sb},(w)}_j],
\end{split}
\end{equation}
and the correlation deficit of the block is
\begin{equation}\label{eq:deficit_block}
  \dFsb(j,t;w)=\Fsb(j,t;w)-\Fs(j,t;w)-\Fb(j,t;w),
\end{equation}
the sensitivity the block carries beyond what its spins and its modes hold separately.
The block deficit $\dFsb(j,t;w)$ [Eq.~\eqref{eq:deficit_block}] is our primary local diagnostic. It measures what a joint reading of the block holds beyond what its spin and bath parties hold separately, mirroring the bipartite Fisher-information surplus~\cite{TothPetz2013,Gessner2016}; being a difference of restricted-access precisions, it is bookkeeping, not itself an attainable resource. Non-superadditivity means $\dFsb$ is not sign-definite; it is positive when the sensitivity resides in spin--bath correlations, and would turn negative in a redundant, Darwinism-like regime in which fragments of either party read the phase independently~\cite{Zurek2009,Brandao2015,Ferte2024}. The probes here do not enter that regime, since for one excitation the whole-chain residual cannot come out negative [App.~\ref{app:method}]; under the Jaynes--Cummings coupling it vanishes identically [Eq.~\eqref{eq:jc_main}], and every block deficit computed here is non-negative. All QFIs are evaluated directly from the tangent state, the $\theta$-derivative of the encoded state at $\theta=0$ [App.~\ref{app:method}].

A coarser summary, running in parallel to the block decomposition, is useful when the probe is spread
along the entire chain. A bar denotes a site-by-site reading of the whole chain. The site sums
$\Fsbar,\Fbbar$ and the whole-chain residual $\dFsbbar$ split the conserved total into spin, bath, and
correlation parts in the same way as the block quantities $\Fs,\Fb,\dFsb$ do for a single block. Summing
the single-site densities,
\begin{equation}\label{eq:sitesum}
  \Fsbar(t)=\sum_{j}\Fs(j,t;1),\qquad
  \Fbbar(t)=\sum_{j}\Fb(j,t;1),
\end{equation}
gives site-resolved figures of merit, each term optimized on its own site; neither is an achievable
precision. A measurement performed site by site and processed jointly also draws on correlations between the
outcomes, so it can exceed $\Fsbar$, while a collective measurement on all spins at once is bounded instead by
the QFI of the full spin marginal. The two references carry no general ordering---in a redundant,
Darwinism-like regime the site sum can exceed any joint value---but for a single excitation the QFI of the
full spin marginal equals $\Fsbar$ exactly at the encoding point [App.~\ref{app:method}], so collective
readout gains nothing there over site-resolved readout. The
residual against the conserved total,
\begin{equation}\label{eq:deficit}
  \dFsbbar(t)\equiv\Fglob-\Fsbar(t)-\Fbbar(t) ,
\end{equation}
defined as this residual and not as a site sum of block deficits, is therefore what site-resolved readout
misses, and it contains coherence between different sites of the chain as well as spin--bath correlation. It
acts as a bookkeeping residual, and we use
it only where the two contributions are distinguishable on other grounds, as in Sec.~\ref{sec:hol}, where the
single excitation carries a cloud whose overlap fixes the suppression directly. The gap between $\Fglob$ and
the summed single-site QFIs diagnoses multipartite coherence, in the spirit of the QFI-based entanglement criteria~\cite{Hyllus2012,Toth2012,PezzeSmerzi2009}, which compare $\FQ$ against the shot-noise value. For one excitation the site sums cannot together
exceed the conserved total [App.~\ref{app:method}], so the residual cannot turn negative.

\section{Single-site encoder}\label{sec:dynamics}
We now follow the local quantum Fisher information after the classical parameter $\theta$ has been written into the chain by a rotation of a single spin at the left edge, with the bath starting in vacuum. The rotation places the chain in a superposition of the vacuum and the one-excitation sector; the tangent state lies in the one-excitation sector, and from that moment the exchange carries the excitation along the chain while the bath acts on it.

The two couplings affect this flow differently. Jaynes--Cummings conserves
the total number of excitations, which confines the dynamics to a single-particle sector and makes the
problem exactly solvable; it also lets the bath spectrum be tuned between full return and permanent loss.
Holstein conserves the number of spin excitations instead, leaving the excitation a spin flip that drags a
cloud of bosons behind it, which puts the problem beyond exact solution but within reach of tensor
networks. Whether a departed share returns is governed by the bath spectrum; commensurate coupled-system frequencies return it periodically, while their spread dephases the channels and prevents full revival (Figs.~\ref{fig:jc} and~\ref{fig:jc_det}).

\subsection{Jaynes--Cummings coupling}\label{sec:jc}
The Jaynes--Cummings coupling~\eqref{eq:jc} converts a spin excitation into a bath boson on the same site and back, conserving the total excitation number $\hat N_{\rm exc}=\sum_i(\nm_i+\ad_i\ah_i)$. Because the spin-excitation amplitude is $\sin(\theta/2)$, the bath state acquires a first-order dependence on $\theta$, giving $\partial_\theta\rhoh^{\rm b}|_{0}\ne0$ and $\Fb>0$, so a measurement confined to the modes can reach the phase. For a single excitation
$\hat N_{\rm exc}=1$ at all times, so the accessible space is the $2L$-dimensional single-particle sector
$\{\ket{j}\ket{\varnothing}\}\cup\{\ket{\FM}\ket{1;j}\}$ and the problem reduces to single-particle
diagonalization. The state is
\begin{equation}
\ket{\Psi(t)}=\sum_j a_j(t)\ket{j}\ket{\varnothing}+\sum_j b_j(t)\ket{\FM}\ket{1;j},
\label{eq:jc_state}
\end{equation}
whose two branches pair the spins with definite, mutually orthogonal bath states. In the spin-excitation branch every bath mode is
in the vacuum $\ket{\varnothing}$, and in the boson branch every spin is up, $\ket{\FM}$. Each branch
therefore pairs the spins with one definite bath state, and tracing the bath out leaves the encoded
single-site state of Eq.~\eqref{eq:a_rho1} with its accompanying overlap equal to unity. What survives is a
spin coherence of magnitude $\tfrac12|\sin\theta|\,|a_j|$, and it is that coherence, not the populations,
that carries the phase: Eq.~\eqref{eq:a_fs} then gives $\Fs(j,t)=|a_j(t)|^2$, and the same construction on
a bath mode gives $\Fb(j,t)=|b_j(t)|^2$. Because each branch pairs the encoded spin with a definite, unentangled bath state, the
joint spin--bath block adds nothing beyond its parts, $\Fsb=\Fs+\Fb$, and the deficit is zero,
\begin{equation}
\begin{split}
&\Fs(j,t)=|a_j(t)|^2,\quad \Fb(j,t)=|b_j(t)|^2,\quad \dFsb(j,t)=0,\\
&\Fsbar=\sum_j|a_j|^2=1-\Fbbar.
\end{split}
\label{eq:jc_main}
\end{equation}
The flow is a pure two-way exchange; $\Fsbar$ is the fraction of the encoded excitation that is still a
spin flip, and the complement has swapped coherently and reversibly into the bath. Under Holstein the
branches do not factorize, and the block QFI is $|\psi_j|^2|\braket{\varnothing}{\eta_j}|^2$
[Eq.~\eqref{eq:a_fs}], an occupation multiplied by a cloud overlap.

\begin{figure}[!tp]
  \centering
  \includegraphics[width=\columnwidth]{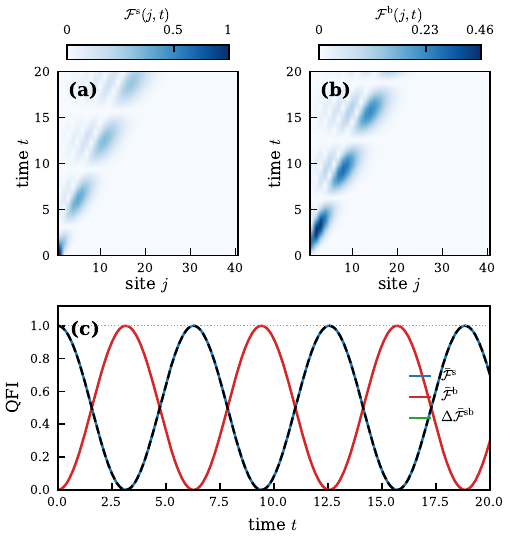}
  \caption{Jaynes--Cummings exchange at resonance on the uniform XX chain [exact single-particle solution,
  $L=40$, $\omega_0=0$, $t_B=0.5J$, $\lambda=0.5J$, edge encoding]. (a),(b)~Spin and bath QFI densities
  $\Fs(j,t)$ and $\Fb(j,t)$. By Eq.~\eqref{eq:jc_main} these equal the occupations exactly,
  and the block deficit vanishes identically.
  (c)~Site-summed shares (dashed: closed-form result).}
  \label{fig:jc}
\end{figure}

\begin{figure}[!tp]
  \centering
  \includegraphics[width=\columnwidth]{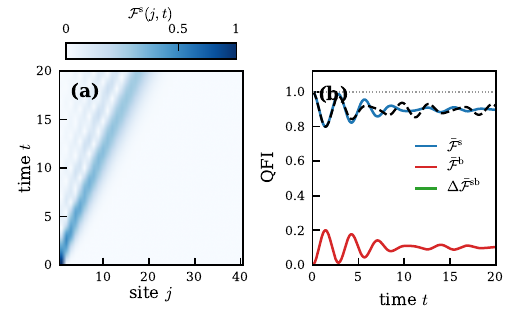}
  \caption{Jaynes--Cummings off resonance [$L=40$, $\omega_0=2J$, $t_B=0.3J$, $\lambda=0.5J$; exact
  single-particle solution, edge encoding]. (a)~Single-qubit QFI density $\Fs(j,t)$; the occupation maps
  coincide with the QFI densities, as in Fig.~\ref{fig:jc}, and are omitted. (b)~Site-summed
  shares (dashed: closed-form result): the site-summed share $\Fsbar$ collapses to a plateau as the
  channels dephase, while the deficit $\dFsbbar$ remains zero.}
  \label{fig:jc_det}
\end{figure}

Written in momentum modes $q$ -- a description exact for a translation-invariant chain and used here as the
infinite-chain reference for the open model of Eq.~\eqref{eq:H} -- the swap splits the dynamics into $L$
independent two-level systems, each pairing a spin excitation and a boson at the same momentum, coupled by
$\lambda$, detuned by $\delta_q=\omega_0+(J-2t_B)\cos q$, and exchanging the excitation at the Rabi
frequency
\begin{equation}
\Omega_q=\sqrt{\delta_q^2+4\lambda^2}.
\label{eq:jc_bands}
\end{equation}
The plotted curves (Figs.~\ref{fig:jc},~\ref{fig:jc_det}) are the exact open-chain diagonalization, which
the momentum forms reproduce until the packet reflects off an edge [App.~\ref{app:jc}]. A local encoding
excites all momenta equally, each $q$ channel oscillates between spin flip and boson, and whether the
leaking sensitivity returns or dephases depends on how widely $\Omega_q$ disperses across the band.

Transforming back to real space for an encoding at site $k$, the amplitudes to be a spin excitation at site $j$ or a
boson at site $j$ are
\begin{subequations}\label{eq:jc_realspace}
\begin{align}
a_j(t)&=\int_{-\pi}^{\pi}\frac{dq}{2\pi}\,e^{iq(j-k)}\,e^{-i\bar\varepsilon_q t}
\Big[\cos\tfrac{\Omega_q t}{2}+i\tfrac{\delta_q}{\Omega_q}\sin\tfrac{\Omega_q t}{2}\Big],\label{eq:jc_aj}\\
b_j(t)&=-i\int_{-\pi}^{\pi}\frac{dq}{2\pi}\,e^{iq(j-k)}\,e^{-i\bar\varepsilon_q t}\,
\tfrac{2\lambda}{\Omega_q}\sin\tfrac{\Omega_q t}{2},\label{eq:jc_bj}
\end{align}
\end{subequations}
with the mean band $\bar\varepsilon_q=\tfrac12(\varepsilon^m_q+\varepsilon^b_q)$, $\varepsilon^m_q=-J\cos q$,
$\varepsilon^b_q=\omega_0-2t_B\cos q$. The space--time densities follow as $\Fs(j,t)=|a_j(t)|^2$,
$\Fb(j,t)=|b_j(t)|^2$, and $\dFsb(j,t)=0$ throughout, even as the state acquires spin--bath
entanglement. At $\lambda=0$ the bracketed Rabi factor in
Eq.~\eqref{eq:jc_aj} reduces to the pure phase $e^{i\delta_q t/2}$, which combines with
$e^{-i\bar\varepsilon_q t}$ into the free spin phase $e^{-i\varepsilon^m_q t}$; the boson channel closes, leaving the free excitation
\begin{equation}
\Fs(j,t)=J_{j-k}(Jt)^2,\qquad \Fb=0,
\label{eq:jc_bessel}
\end{equation}
the Bessel light cone of the bare XX chain, along which the encoded QFI spreads without loss. Switching on
the coupling pulls the modulus of the bracketed factor below unity, modulating this profile, and opens the bath channel $\Fb(j,t)$, which carries
the swapped share and, at resonance, returns it to the spins one period later (Fig.~\ref{fig:jc}). A rise $d\Fsbar/dt>0$ of the site-summed spin share signals sensitivity returning from the bath, in analogy with Fisher-information backflow~\cite{Lu2010,Rivas2014,Breuer2016,Chruscinski2014}, although $\Fsbar$, being a sum of marginal QFIs, is not itself a non-Markovianity measure. The leading decay
$\Fsbar=1-\lambda^2t^2+O(t^4)$ is bath-independent, set by the bare swap, and the bath enters only at $O(t^4)$.

When the boson and spin bands are parallel ($t_B=J/2$), the momentum-independent detuning $\delta_q=\omega_0$ synchronizes all channels to a single frequency $\Omega=\sqrt{\omega_0^2+4\lambda^2}$. The spin--boson swap factorizes from the Bessel propagation [App.~\ref{app:jc}], so the QFI oscillates fully between the parties without dephasing. Closing the gap ($\omega_0=0$) sets the swap depth to unity. The encoded QFI spreads along the shared light cone while transferring completely back and forth, $\Fsbar=\cos^2(\lambda t)$.

When the bands differ, the frequencies disperse and the channels dephase [App.~\ref{app:jc}]. The spin excitation and swapped boson drift apart at different velocities, preventing recombination. The site-summed share $\Fsbar$ thus collapses to a plateau (Fig.~\ref{fig:jc_det}); the plateau is the off-resonant limit of the fractional decay familiar from emitter--photon bound states in structured reservoirs~\cite{JohnWang1990,GonzalezTudela2017,GonzalezTudela2017b}, the parameters of Fig.~\ref{fig:jc_det} lying in the far-detuned regime. Tuning $t_B$ thus interpolates, over the accessible window, between reversible memory and irreversible loss. The deficit remains zero, and the lost QFI dephases across the bath modes.

Conversely, a large bath gap detunes the swap and restricts the amplitude to $4\lambda^2/\omega_0^2$; without resonant bath states the excitation is dressed but not absorbed, and $\Fsbar$ stays close to unity.

\subsection{Holstein coupling}\label{sec:hol}
The Holstein coupling~\eqref{eq:hol} reaches the bath only through the excitation density $\nm_i$. Because the density is even in $\theta$, rotating the probe by $\theta$ or $-\theta$ creates the same bath displacement. The bath marginal therefore has no term linear in $\theta$, giving $\partial_\theta\rhoh^{\rm b}|_{0}=0$ and $\Fb=0$ at $\theta=0$. We refer to this symmetry constraint on the bath marginal as the destination rule. The bath marginal, a classical mixture of the vacuum and the cloud-bearing branch, changes rank at $\theta=0$, and the QFI of a rank-changing family is discontinuous there~\cite{Safranek2017,Seveso2020}. At any $0<\theta<\pi$ the mixing weights $\cos^2(\theta/2)$ and $\sin^2(\theta/2)$ carry Fisher information about $|\theta|$, and the numerical scan of App.~\ref{app:theta} finds a bath QFI of order the correlation deficit, $\simeq1-e^{-\Nbtot}$, already at the smallest nonzero angle computed, while under Jaynes--Cummings it is smooth throughout. The symmetry forbids only the first-order term at $\theta=0$; the bath registers the magnitude of the rotation, never its sign. This first-order null is protected by parity at any temperature, though the discontinuity at $\theta=0$ is a zero-temperature feature, the marginal being full rank once the bath is thermal. At $\theta=0$ a measurement of the modes alone returns nothing about the encoded phase to first order, and the sensitivity the spins lose is carried, in the tangent state, by the correlation between them and the bath.

Because the Holstein coupling displaces the bath oscillator while conserving the excitation number, the excitation remains a spin flip and the bath registers only its position---a spatially resolved independent-boson, pure-dephasing model~\cite{Palma1996,Leggett1987}.

\begin{figure}[tp]
  \centering
  \includegraphics[width=\columnwidth]{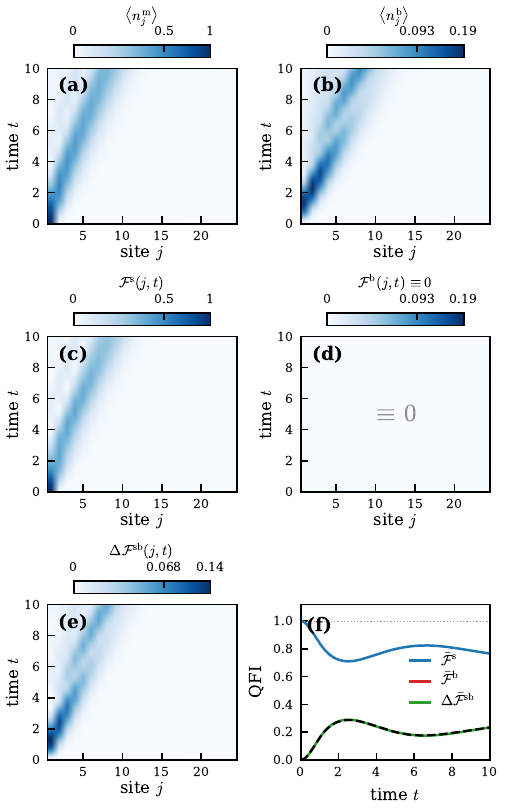}
  \caption{Holstein coupling on the uniform XX chain [$L=24$, $\omega_0=2J$, $t_B=0.5J$, $\lambda=0.5J$, edge encoding; TEBD]. (a),(b)~Occupations on the tangent state. (c),(d)~QFI densities $\Fs$ and $\Fb\equiv0$; the boson cloud in (b) is populated while (d) stays empty of phase information, and the color scale of the identically zero panel is nominal. (e)~Block deficit $\dFsb$. (f)~Whole-chain residual $\dFsbbar$ tracking $1-e^{-\Nbtot}$ (dashed).}
  \label{fig:hol}
\end{figure}

A coherent boson cloud forms under each excitation and co-moves with it as it hops [Fig.~\ref{fig:hol}(b)]. Because the bath displacement is conditioned on the excitation position, tracing out the bath suppresses the spin coherence between different site amplitudes.

The recoverable $\Fsbar$ falls below the conserved $\Fglob=1$ by an amount estimated for a coherent cloud as follows. An excitation at site $j$ is dressed by its coherent cloud $\ket{\eta_j}$; tracing out the bath suppresses the spin coherence by the overlap with the reference vacuum, $|\braket{\varnothing}{\eta_j}|^2=e^{-\bar n_j}$ [App.~\ref{app:method}]. Since the cloud co-moves with the excitation ($\bar n_j\simeq\Nbtot$), summing over the excitation distribution yields
\begin{equation}\label{eq:hol_deficit}
\Fsbar\simeq e^{-\Nbtot},\qquad \Fbbar=0,\qquad \dFsbbar\simeq1-e^{-\Nbtot},
\end{equation}
with $\Nbtot=\sum_l\langle n^{\rm b}_l\rangle$ the total boson number. Each site's spin share is suppressed by its cloud overlap, and the lost sensitivity is held between the excitation and the bath it dragged.

\begin{figure}[tp]
  \centering
  \includegraphics[width=\columnwidth]{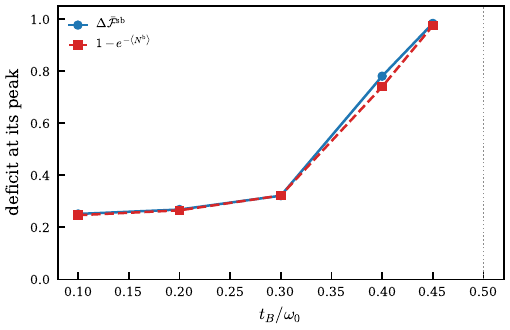}
  \caption{Correlation deficit of the itinerant Holstein coupling as the bath band softens [$L=20$, $\omega_0=2J$, $\lambda=0.5J$; TEBD]. Deficit at its peak time (circles) and $1-e^{-\Nbtot}$ (squares) vs bath hopping $t_B/\omega_0$ (dotted: soft-mode boundary $t_B=\omega_0/2$).}
  \label{fig:softmode}
\end{figure}

A conditionally displaced bath suppresses the spin coherence by $e^{-\bar n/2}$, and the QFI---quadratic in the coherence---by the zero-temperature independent-boson dephasing factor $e^{-\bar n}$~\cite{Duke1965,Mahan2000,Palma1996}. This exponential scaling is exact for a static excitation [App.~\ref{app:ghz_exact}], and persists once the excitation is free to carry its cloud [Fig.~\ref{fig:hol}(f)], with the deficit tracking the boson ring-down while $\Fbbar$ stays at zero. While the itinerant test operates at modest occupations ($\Nbtot\lesssim0.34$), pushing $t_B$ toward the soft-mode limit ($\omega_0/2$) drives $\Nbtot$ above unity, where the simulated deficit follows the exponential and departs from linear response [Fig.~\ref{fig:softmode}].

The correlation deficit is largest when the bath features a soft mode. A localized excitation drives each bath mode to an occupation oscillating as $\sin^2(\omega_q t/2)$, returning the sensitivity periodically. As $t_B$ approaches $\omega_0/2$, the lowest frequency $\omega_q \to 0$ and its occupation grows quadratically in time, as for a resonantly driven oscillator. This mode is driven resonantly and never empties, so the sensitivity it absorbs does not return. The correlation deficit therefore climbs toward its maximum as $t_B \to \omega_0/2$. Although this divergence is a thermodynamic-limit statement---the lowest mode frequency of a finite chain remains strictly positive---approaching this limit demands rapidly growing bond dimensions and boson cutoffs as the soft mode delocalizes [App.~\ref{app:convergence}]. In the stable regime ($t_B<\omega_0/2$) the correlation deficit rises from $\dFsbbar\simeq0.25$ at $t_B/\omega_0=0.1$ toward unity as the soft mode develops, and tracks the $1-e^{-\Nbtot}$ prediction up to $\Nbtot\simeq3.7$. In the dispersive regime it runs a few percent above the prediction [App.~\ref{app:method}], because the itinerant cloud, spread across several bath sites, is no longer a product of on-site coherent states. The mode--mode correlations built up in transit lower the joint vacuum overlap below the product estimate.

\section{GHZ encoder on the full chain}\label{sec:ghz}
A single excitation carries a sensitivity bounded by $\Fglob=1$, so the flow followed so far redistributes a shot-noise resource. We now evaluate the decoherence cost imposed by the bath on a macroscopic probe.
We encode the phase on a GHZ state of $n$ spins, $\Fglob=n^2$, and take $L=n$ so that every spin sits on its own bath mode and the initial probe gives the exchange nothing to transport. This isolates the action of the bath on an entangled multiparticle probe, and explores whether the destination and the return established for one excitation still hold when the sensitivity is stored in $n$-body correlations, and how fast the advantage is lost as $n$ grows.

The probe is a Greenberger--Horne--Zeilinger state on the $n$ spins, carrying the Heisenberg value
$\Fglob=n^2$ in $n$-body correlations that a single excitation cannot form, with the bath in vacuum,
\begin{equation}
  \ket{\Psi(0)}=\tfrac1{\sqrt2}\big(\ket{\uparrow}^{\otimes n}+\ket{\downarrow}^{\otimes n}\big)
                \otimes\ket{\varnothing},
\label{eq:ghz_state}
\end{equation}
with the phase imprinted through $\hat G=\sum_{i=1}^{n}\Sz_i$, the natural encoding axis for a GHZ probe. Because the enhancement lives in $n$-body
correlations, tracing to any $w<n$ spins gives $\Fs(j,t;w)=0$ identically, for all $t$ and under either
coupling. The signal appears only at the full block, and the gap between $w<n$ and $w=n$ is the correlation
deficit of Sec.~\ref{sec:model} carrying all of it.

At $L=n$ the exchange annihilates both branches of the initial state,
$\Hh_{\rm s}\ket{\uparrow}^{\otimes n}=\Hh_{\rm s}\ket{\downarrow}^{\otimes n}=0$. Under the Holstein
coupling each branch stays a polarization eigenstate at all times, so the exchange never acts and the
dynamics is that of the bath alone, and for an Einstein bath ($t_B=0$, every mode at the bare frequency
$\omega_0$) the readable share has a closed form, exact at
every $n$ with no boson cutoff [App.~\ref{app:ghz_exact}],
\begin{equation}
  \Fs(1,t;n)/n^2=\big[e^{-2g(1-\cos\omega_0 t)}\big]^{n},\qquad g=(\lambda/\omega_0)^2 .
  \label{eq:ghz_hol_frozen}
\end{equation}
This is a single-site return factor raised to the $n$-th power, the exact recoherence of a qubit register
dephased by oscillators~\cite{Unruh1995}. Each qubit meets its own mode, so the
coherence decays $n$ times as fast as one qubit's~\cite{Huelga1997,Demkowicz2012,Frowis2011}, the independent-bath
fragility of a GHZ probe, linear in $n$ and distinct from the $n^2$ superdecoherence a common bath would
produce~\cite{Palma1996,Monz2011,Reina2002}.

The Jaynes--Cummings coupling forfeits that generality, since it raises a spin in the lower branch that the
exchange then transports, and a closed form survives only in the frozen limit $J=t_B=0$, where it is exact,
\begin{equation}
\begin{gathered}
\Fs(1,t;n)=\frac{2n^2x}{1+x+y},\qquad x=r^{\,n},\qquad y=(1-r)^{n},\\
r=1-\frac{4\lambda^2}{\Omega_0^2}\sin^2\tfrac{\Omega_0 t}{2},\qquad
\Omega_0=\sqrt{\omega_0^2+4\lambda^2},
\end{gathered}
\label{eq:ghz_frozen}
\end{equation}
with $r$ the single-site probability that an excitation has not converted to a boson. The denominator counts the weight the two spin configurations carry. The coupling
annihilates $\ket{\uparrow}^{\otimes n}\ket{\varnothing}$, so that branch holds $\tfrac12$ throughout; the
lower branch keeps $\tfrac12x$ with its bath still in vacuum, and once all $n$ excitations have converted it
returns weight $\tfrac12y$ to the configuration $\ket{\uparrow}^{\otimes n}$, now paired with an
$n$-boson state and so carrying population without coherence. Both $x$ and $y$ shrink the readable share;
$x$ oscillates back to unity and restores it, while $y$ never does. Spin-excitation-number conservation removes the whole effect, since no population is
transferred and both branch weights stay at $\tfrac12$ over a common bath reference; Eq.~\eqref{eq:ghz_hol_frozen}
therefore carries no such denominator and holds at any $J$.

Splitting the $w=n$ block into its three shares (Fig.~\ref{fig:ghz_dynamics}, top row) shows the destination rule surviving the
change of probe. The spin site sums of Eq.~\eqref{eq:sitesum} vanish identically here; the bath site sums
vanish identically under Holstein and to numerical precision under Jaynes--Cummings, where only an
exponentially small amplitude for all $n$ bosons to bunch on a single mode survives the trace. The joint
block is the whole pure state, so the budget closes on the conserved total. Holstein keeps the bath empty of phase and
drives the entire signal into the block deficit, which swings up to $n^2$ and back while the bath marginal
carries none of it; Jaynes--Cummings does populate the bath share, though it remains well below the deficit, so the larger part sits where neither marginal reaches it.

The destination rule therefore holds at any excitation number, but it does not extend to the deficit. With one excitation the Jaynes--Cummings state has two branches, each pairing the spins with a single definite
bath state; the marginals jointly exhaust the phase information and the QFI deficit vanishes [App.~\ref{app:method}]. With $n$ excitations
the swap acts on every site independently, the state carries every configuration of swapped and unswapped
sites at once, and the bath thereby registers which spins still hold theirs; spins and bath are then correlated whatever
the coupling does with the phase, so the deficit vanishes only in the single-excitation case.

\subsection{Scaling limits of the metrological advantage}
\begin{figure}[tp]
  \centering
  \includegraphics[width=\columnwidth]{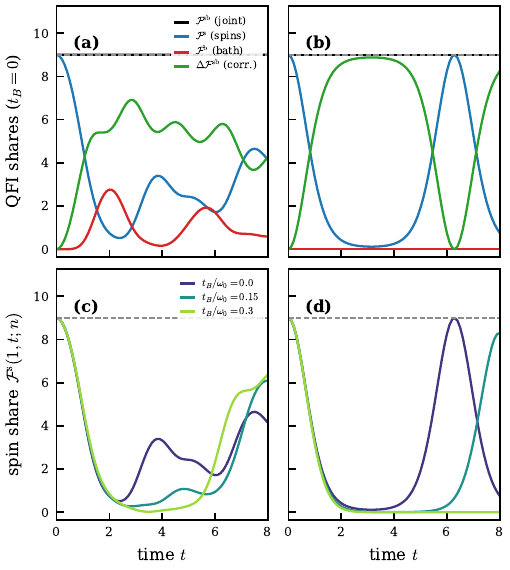}
  \caption{GHZ probe dynamics without transport [$L=n=3$, $\omega_0=J$, $\lambda=0.6J$; evolved under the full
  Hamiltonian; the exchange annihilates the initial branches and never acts under Holstein, while under
  Jaynes--Cummings it transports the spin raised in the lower branch]. Top row: three-share budget at $t_B=0$ for (a)~Jaynes--Cummings and (b)~Holstein, summing to the conserved $\Fglob=n^2=9$ (black). Holstein maintains $\Fb\equiv0$, trapping the lost sensitivity within $\dFsb$. Bottom row: recoverable spin-block QFI $\Fs(1,t;n)$ at $t_B/\omega_0=0,\,0.15,\,0.3$ for (c)~Jaynes--Cummings, which retains a partial revival at every $t_B$, and (d)~Holstein, which revives fully at $t_B=0$ ($t=2\pi/\omega_0$), is delayed at $t_B/\omega_0=0.15$, and stays suppressed at $0.3$; dashed line $n^2$.}
  \label{fig:ghz_dynamics}
\end{figure}

Figure~\ref{fig:ghz_dynamics} (bottom row) tracks the recoverable share $\Fs(1,t;n)$ as the bath modes are coupled and $t_B$
turns on. Jaynes--Cummings brings back a partial revival at every $t_B$, the exchange with the
bath continuing whatever the modes do, though the return is incomplete once the deficit has grown. The Holstein revival is
far more sensitive to $t_B$. At $t_B=0$ the single mode is a perfect memory and $\Fs$ returns in full at
$t=2\pi/\omega_0$; a small $t_B$ detunes the modes from one another and delays the revival; by $t_B/\omega_0=0.3$ the
revival has left the window altogether and the whole enhancement remains in the correlation deficit. For the Holstein coupling the advantage is unrecoverable once the bath disperses.

\begin{figure}[tp]
  \centering
  \includegraphics[width=\columnwidth]{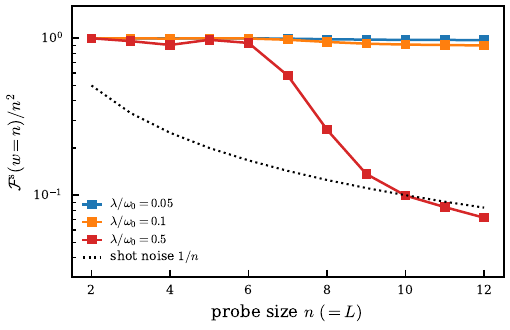}
  \caption{Scaling of the Heisenberg advantage with probe size for the frozen Holstein GHZ probe on the
  dispersive bath [$L=n$, $t_B/\omega_0=0.3$; the exchange never acts on the Holstein branches; exact closed form of
  App.~\ref{app:ghz_exact}]. Best revival of the readable share $\Fs(w{=}n)/n^2$ against probe size $n$
  for $\lambda/\omega_0=0.05,\,0.1,\,0.5$; dotted curve the shot-noise value $1/n$. On the Einstein bath
  ($t_B=0$, not shown) the best revival returns to unity at every $n$ and every coupling, and only the
  anti-revival minima deepen, as $e^{-4ng}$ [Eq.~\eqref{eq:ghz_hol_frozen}]. With dispersion the mode
  returns stop coinciding and the best revival falls with $n$, dropping below $1/n$ by $n=12$ at
  $\lambda/\omega_0=0.5$, where the advantage is lost outright.}
  \label{fig:ghz_nL}
\end{figure}

On the Einstein bath the scaling is exact for every $n$ [Eq.~\eqref{eq:ghz_hol_frozen}]. The exponent
vanishes at $t=2\pi k/\omega_0$ for any $n$, so increasing $n$ deepens the
minima, as $e^{-4ng}$, without lowering the revival height, and the readable window narrows. The frozen probe of App.~\ref{app:ghz_exact} is
solved in closed form for every $n$ from the $n\times n$ bath matrix, with no cutoff. Dispersion closes the
window altogether, because the modes then no longer share a period and cannot return the share together;
the more spins the probe holds, the more returns must coincide for the coherence to rebuild. The best revival therefore falls as the probe grows, and by a dozen spins at the strongest coupling shown
it drops below the ideal shot-noise level $1/n$ of $n$ noiseless independent spins (Fig.~\ref{fig:ghz_nL}). The Heisenberg advantage is
then lost, with no revival at any time examined; unlike the Einstein-bath case, the share held in the
correlations is never returned to the spins.
Jaynes--Cummings degrades already at $t_B=0$, through the same lower-branch spin flip that
spin-excitation-number conservation forbids under Holstein.

\section{GHZ encoder on a perfect-state-transfer chain}\label{sec:pst}
Carrying the Heisenberg advantage across the bath requires the probe to travel, and away from $L=n$ the exchange
carries the probe's own support out of the $w=n$ block within a few hops, at the same rate whether the bath is
coupled or not, so on a uniform chain transport alone destroys the probe before the bath acts. An engineered chain removes this loss and carries the block to a distant read-out, where any remaining
loss is attributable to the bath.
Carrying a probe from where it is prepared to where it is read is the setting of distributed and networked
sensing~\cite{Proctor2018,Zhuang2018,Guo2020,Malia2022}, with a spin chain as the channel. An unmodulated
chain transfers a state with a fidelity that decays with its length~\cite{Bose2003}. With couplings $J_i\propto\sqrt{i(L-i)}$ the single-particle spectrum is equally spaced and the propagator
mirrors the register at $t_\star$~\cite{Christandl2004}, in every excitation sector at
once~\cite{Albanese2004,Kay2010}, so a block on sites $1\ldots n$ arrives at the far end intact.
Transfer of an entangled multi-qubit register on such a closed chain reaches fidelities
above $0.99$ for up to three qubits~\cite{Michel2026}. The profile
makes the single-excitation block of the Hamiltonian the $\hat J_x$ operator of a spin $j=(L-1)/2$, so the single-particle propagator is a
Wigner rotation matrix and the mirror time follows in closed form, $t_\star=\pi L/2J$ [App.~\ref{app:jc}]. For the matched comparison of the two couplings we take $L=8$ and $n=3$, launch the probe on sites
$1\ldots3$, and read the $w=n$ block on the mirrored sites $L\!-\!2\ldots L$, so that $t_\star=4\pi$ for our
normalization $\max_i J_i=J$; the Holstein case is extended to $L=20$--$30$ below.

Switching the bath off gives a control the uniform chain cannot provide, since the block then arrives at
the far end with its Heisenberg advantage intact and every loss reported below is attributable to the bath
alone. A static bath ($t_B=0$) is known to modulate the arrival without preventing
transfer~\cite{Burgarth2006,Baum2022}, the Jaynes--Cummings coupling then decoupling into $2\times2$ blocks
[App.~\ref{app:jc}]; a mobile bath ($t_B\neq0$)
breaks that decoupling, because the boson hopping no longer commutes with the engineered exchange, and what
it costs the probe is not fixed by the static solution but depends on which coupling carries it
[App.~\ref{app:jc}].

\subsection{Block QFI versus register fidelity}\label{sec:fidelity}
A transfer fidelity certifies the arrival of the state; a metrological readout needs the sensitivity the
arrived state can still deliver. Both depend on the surviving coherence, through different powers of it.

Let us write the reduced state of the arriving $w=n$ spin block in the basis that carries the probe,
\begin{equation}\label{eq:ghz_sector}
\begin{gathered}
  p_{\rm d}=\bra{\downarrow^n}\rhoh^{{\rm s},(n)}\ket{\downarrow^n},\qquad
  p_{\rm u}=\bra{\uparrow^n}\rhoh^{{\rm s},(n)}\ket{\uparrow^n},\\
  c=\big|\bra{\downarrow^n}\rhoh^{{\rm s},(n)}\ket{\uparrow^n}\big| .
\end{gathered}
\end{equation}
Both states are eigenstates of $\hat G=\sum_i\Sz_i$ with eigenvalues $\mp n/2$, so the encoding turns the
coherence and leaves the populations alone. The fidelity against the target register and the block QFI then
follow in closed form [App.~\ref{app:fidelity}],
\begin{equation}\label{eq:fid_vs_qfi}
  F=\frac{p_{\rm u}+p_{\rm d}}{2}+c,\qquad
  \frac{\Fs(1,t;n)}{n^2}=\frac{4c^{2}}{p_{\rm u}+p_{\rm d}} ,
\end{equation}
the first maximized over the relative phase of the two branches, which a known transfer phase corrects
locally; the fidelities reported below are these phase-calibrated values. Both expressions are exact within
the GHZ sector of the block, and the residual against the QFI of the full $2^n$-dimensional block stays
below $2\times10^{-5}$ here [App.~\ref{app:fidelity}]. They agree whenever the
weight the block retains is a pure pair of GHZ branches, both reaching unity only for perfect delivery, and
they separate once the channel costs coherence faster than it costs weight.

The two branches are not equivalent under this Hamiltonian. The $p_{\rm u}$ branch ($\ket{\uparrow^n}$, no spin excitations) is annihilated by the exchange and left undisplaced by the Holstein
coupling, which acts only through the excitation density; its weight is conserved at $\tfrac12$. The $p_{\rm d}$ branch ($\ket{\downarrow^n}$, $n$ spin excitations) must cross the chain, each excitation dragging the cloud that dephases it, so it loses weight to sites outside the block and coherence to the bath simultaneously. A longer chain lengthens both effects, and by $L=20$ the $p_{\rm d}$ branch retains less than half the weight of $p_{\rm u}$ ($0.21$ against $0.50$ at $t_\star$ on the static bath).

Both quantities are then read off the same shrinking coherence, and Eq.~\eqref{eq:fid_vs_qfi} converts it
through different powers. The fidelity picks up one power of $c$ and half the surviving weight, so a register that arrives
with its populations intact still scores well while its coherence decays; the QFI picks up $c^2$ and pays
twice over, consistent with the general quadratic relation between fidelity witnesses and the
QFI~\cite{Apellaniz2017,Frowis2011}. The metrological loss is therefore larger than the transfer fidelity indicates. On the static bath the
fidelity exceeds the readable share by $8\%$ of the latter at $L=8$ and by $24\%$ at $L=20$, the gap widening
monotonically with the distance traveled; once the bath disperses it is larger at every length but no longer
monotone in $L$.

\begin{figure}[tp]
  \centering
  \includegraphics[width=\columnwidth]{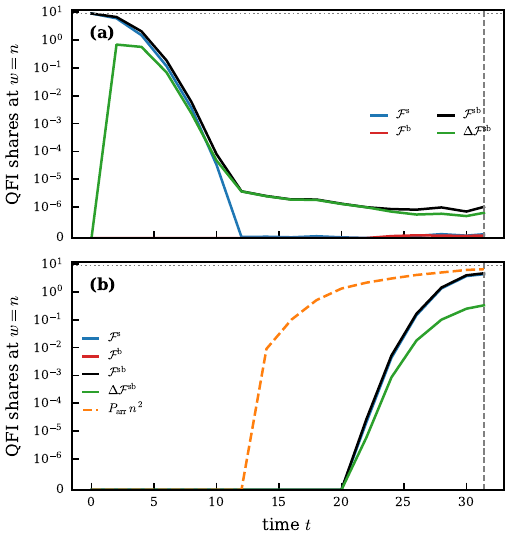}
  \caption{Holstein GHZ block ($n=3$) transported on an engineered $L=20$ chain [$\omega_0=J$,
  $\lambda/\omega_0=0.1$, $t_B/\omega_0=0.3$; TEBD, $N_{\max}=4$]. QFI shares at the source block ($j=1$)
  and the mirror block ($j=L-n+1$) on a symlog scale with linear threshold $10^{-6}$, which is also the
  resolution floor of the late-time tails in (a). $\Fb$ stays at or below $10^{-7}$ throughout, consistent
  with its exact zero; $\Fs$ rebuilds at the mirror to $4.3$ out of $n^2=9$ while $\dFsb$ grows in transit.
  Dashed orange: scaled arrival probability $n^2 P_{\rm arr}$; dashed vertical line $t_\star$, dotted $n^2$.}
  \label{fig:pst_holstein_L20}
\end{figure}

Counting the arriving excitations does worse still, since a population fraction never sees $c$ at all and
scores a register as delivered whenever the excitations are in the right place, whatever phase relation
they retain. Part of that gap comes from weight loss alone, since $n$ excitations arriving independently
already carry the coherence as a product and give, by Eq.~\eqref{eq:fid_vs_qfi},
$\Fs/n^2\to2P_{\rm arr}^{\,n}/(1+P_{\rm arr}^{\,n})$, with $P_{\rm arr}$ the
probability that a launched excitation has reached the mirror block (the dashed curve of
Fig.~\ref{fig:pst_holstein_L20} shows the population count $n^2P_{\rm arr}$ itself); the rest is the
dephasing described above.

\begin{figure}[tp]
  \centering
  \includegraphics[width=\columnwidth]{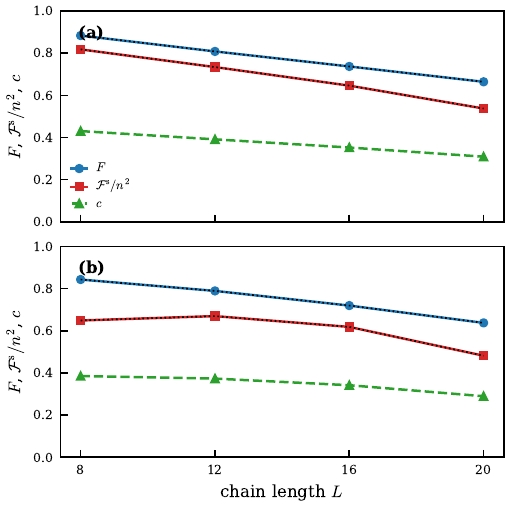}
  \caption{Register fidelity $F$ and readable share $\Fs/n^2$ of an arriving Holstein GHZ probe ($n=3$) vs chain length $L$ ($\lambda/\omega_0=0.1$). (a)~$t_B=0$; (b)~$t_B/\omega_0=0.3$. Circles: $F$; squares: $\Fs/n^2$; triangles: coherence $c$. Dotted: the closed forms of Eq.~\eqref{eq:fid_vs_qfi} evaluated on the simulated $p_{\rm u}$, $p_{\rm d}$, $c$.}
  \label{fig:regfid}
\end{figure}

\subsection{Symmetry constraints under transport}
\begin{figure}[tp]
  \centering
  \includegraphics[width=\columnwidth]{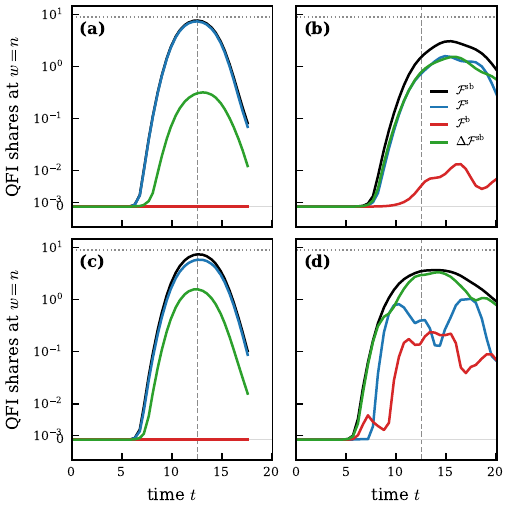}
  \caption{QFI shares of an arriving GHZ block ($n=3$, $L=8$, $\omega_0=J$) on an engineered chain, read at
  the mirrored block ($t_\star$ dashed, $n^2$ dotted; symlog scale). Columns: Holstein at
  $\lambda=0.1\,\omega_0$ (a),(c), holding $\Fb\equiv0$; Jaynes--Cummings at $\lambda=0.5\,\omega_0$
  (b),(d), lifting $\Fb$ to $0.01$--$0.24$. Rows: $t_B=0$ (a),(b) and $t_B/\omega_0=0.3$ (c),(d). The two
  couplings are shown at different strengths because the destination rule is qualitative and
  strength-independent; each is displayed where its mechanism is visible and converged
  (Holstein at $0.5\,\omega_0$ saturates the bond dimension, Jaynes--Cummings at $0.1\,\omega_0$ has
  $\Fb\sim10^{-3}$, invisible on this scale).}
  \label{fig:pst_shares}
\end{figure}

The destination rule of Sec.~\ref{sec:hol} makes no reference to where the excitations are, so a probe
that has crossed five sites and dragged a cloud the whole way is bound by it exactly as the probe at the
encoding site is. The bath marginal stays
empty of phase under the Holstein coupling and acquires it under the Jaynes--Cummings coupling, and the
transit does not weaken the rule (Fig.~\ref{fig:pst_shares}). The share the coupling withholds from the bath is
not returned to the spins either; it accumulates in the correlation deficit, which grows as the block crosses
and grows further when the bath disperses and carries the correlation away with it.

Two channels drain the arriving block, and away from $L=n$ they are distinguishable because the joint block is
now a subsystem, so $\Fsb$ measures on its own how much sensitivity is
present at the block in any form. Sensitivity that arrived but sits in spin--bath correlation can still be recovered by a joint measurement
of the arriving spins and their modes. Sensitivity that never arrived cannot, since it dispersed along the
chain and across the modes while the probe crossed. Only the first channel exists for the
frozen probe of Sec.~\ref{sec:ghz}, where the deficit was the entire loss; under transport it is the
recoverable part of a larger one.

\begin{figure}[tp]
  \centering
  \includegraphics[width=\columnwidth]{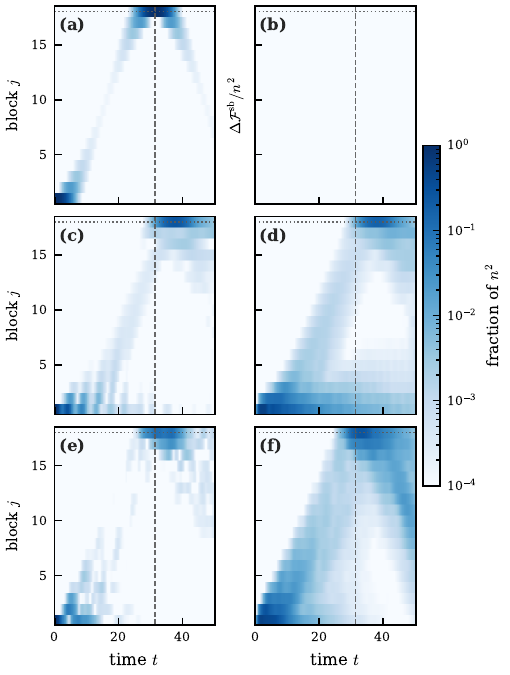}
  \caption{Space-time maps of the spin share $\Fs$ (left column) and the correlation deficit $\dFsb$ (right column), each over $n^2$, for a Jaynes--Cummings GHZ block ($n=3$, $L=20$, $\omega_0=J$) crossing the engineered chain [exact diagonalization, log color scale; dashed line $t_\star$, dotted line the mirror block $j=L-n+1$]. Rows: $\lambda=0$ control (a),(b); $\lambda/\omega_0=0.5$ at $t_B=0$ (c),(d) and at $t_B/\omega_0=0.3$ (e),(f). The control block travels and mirrors with zero deficit (b); under the coupling the deficit forms and remains at the launch site, with a second, weaker episode at the mirror as the block arrives (d), and the mobile bath spreads that weight over mid-chain blocks (f).}
  \label{fig:pst_heatmap}
\end{figure}

The two shares also move differently, which a space-time map separates (Fig.~\ref{fig:pst_heatmap}). The
spin-readable share follows the probe, sitting at the Heisenberg value where the block is launched and
rebuilding at the mirror, but collapsing while the block is in flight, when no contiguous $n$-site window
holds the $n$-body coherence. Even at the best-placed block at each instant, it falls more than two orders
of magnitude below its launched value during the crossing.

The deficit does not propagate; it forms at the encoding site, decays there, and forms again, smaller, at
the mirror as the block arrives, so the map shows two separate events. Dispersing the bath moves that weight downstream earlier and
spreads it over blocks in mid-chain instead of concentrating it at the mirror.

The Holstein coupling reaches the longest chains we compute ($L=20$ for the full QFI shares,
$L=30$ for the transport itself, certified in App.~\ref{app:convergence}), and on the static bath the separation between what
travels and what survives grows monotonically with chain length up to $L=20$. The excitation arrives; with the bath dispersed, the block that receives it retains less than half
the Heisenberg value it was launched with [Fig.~\ref{fig:pst_holstein_L20}], and the bath marginal holds no first-order phase information at any point of the crossing.
The lost sensitivity went into spin--bath correlation beyond the reach of either party alone, and most of the
shortfall comes from coherence lost in transit
[Fig.~\ref{fig:pst_holstein_L20}].

\section{Discussion and Conclusions}\label{sec:concl}
A parameter encoded in a quantum spin chain loses local sensitivity when the spins couple to a bosonic bath. By treating the bath as an explicit quantum many-body system and studying its internal degrees of freedom, we follow the complete time evolution of the joint system. Mapping the quantum Fisher information of every accessible block then reveals where the departed sensitivity goes and whether a restricted measurement can still recover it.

We show that the symmetry of the spin--bath coupling dictates the destination of the information flow. The Jaynes--Cummings interaction routes the sensitivity into the bath marginal, allowing a bath-only readout to extract the encoded phase. The Holstein interaction withholds this information from the marginal, leaving a bath-only readout blind to the encoded rotation.

For entangled multiparticle probes, the coupling symmetry further determines whether the sensitivity revives after initial decay, while soft bath modes act as a permanent information sink. Under quantum transport, we demonstrate that high state fidelity does not guarantee high transferred Fisher information, as the surviving coherence enters the precision limit quadratically rather than linearly.

The correlation deficit quantifies the fraction of the encoded sensitivity that survives exclusively in the joint spin--bath correlations. Tracking this distribution clarifies the precision limits available under restricted measurements.

\begin{acknowledgments}
M.P. acknowledges RES resources provided by Barcelona Supercomputing Center in MareNostrum~5 to
NNO-2025-3-0004.
\end{acknowledgments}

{\it Data availability.---}The data that support the findings of this article are
available from the author upon reasonable request.

\bibliographystyle{apsrev4-2}
\bibliography{refs}

\appendix
\onecolumngrid

\section{Reference state, block reduced states, and the block QFI}\label{app:method}
Because $\Hh$ is independent of $\theta$, the parameter derivative commutes with the evolution and the QFI at
$\theta=0$ is fixed by the reference state and its tangent. Using $-i\Sy_k\ket{\uparrow}_k=\tfrac12\ket{\downarrow}_k$,
\begin{align}
\ket{\Psi_0(t)}&=e^{-i\Hh t}\ket{\FM}\ket{\varnothing},\label{eq:a_ref}\\
\ket{\Theta(t)}&=\partial_\theta\ket{\Psi(\theta,t)}\big|_{0}=\tfrac12\,e^{-i\Hh t}\ket{k}\ket{\varnothing}
=\tfrac12\Big(\sum_{j=1}^{L}\psi_j\ket{j}\ket{\eta_j}+\ket{\FM}\ket{\zeta}\Big),\label{eq:a_tan}
\end{align}
with $\ket{\FM}$ the excitation vacuum, $\ket{\varnothing}$ the bath vacuum, and $\ket{k}=\Sm_k\ket{\FM}$ the
excitation created at the encoding site $k$; the reference $\ket{\Psi_0}$ is stationary. The injected excitation is
either a spin excitation at some site $j$, dressed by
a bath state $\ket{\eta_j}$, or, when the coupling changes the excitation number, transferred to the bath, leaving
the excitation vacuum $\ket{\FM}$ and a bath state $\ket{\zeta}$. Projecting onto the excitation-position states
$\ket{j}=\Sm_j\ket{\FM}$ defines the excitation amplitude and its accompanying bath state,
\begin{equation}
\psi_j(t)\,\ket{\eta_j(t)}=\bra{j}\,e^{-i\Hh t}\,\ket{k}\ket{\varnothing},
\qquad \braket{\eta_j}{\eta_j}=1,
\label{eq:a_eta}
\end{equation}
so $|\psi_j|^2$ is the probability that the excitation is a spin excitation at $j$. A coupling that conserves the number of spin excitations keeps
$\ket{\zeta}=0$ and $\sum_j|\psi_j|^2=1$; one that exchanges them with the bath populates it. For Holstein $\ket{\eta_j}$ is the
boson cloud dressing the spin excitation. For Jaynes--Cummings the spin excitation carries the bare vacuum,
$\ket{\eta_j}=\ket{\varnothing}$ and $\psi_j=a_j$, and $\ket{\FM}\ket{\zeta}=\sum_l b_l\ket{\FM}\ket{1;l}$ is the
single swapped boson [App.~\ref{app:jc}], so $\Fs(j)=|a_j|^2$ and $\Fb(l)=|b_l|^2$. The reduced state of a
contiguous $w$-spin block $B^{(w)}_j$ is obtained by tracing out the bath and every spin outside the block,
\begin{equation}
\rhoh^{(w)}_j(\theta,t)=\Tr_{\rm bath}\,\Tr_{\overline{B^{(w)}_j}}\,\dyad{\Psi(\theta,t)},\qquad
\overline{B^{(w)}_j}=\{1,\dots,L\}\setminus B^{(w)}_j,\label{eq:a_rdm}
\end{equation}
and its QFI is the spectral form
\begin{equation}
\FQ[\rhoh]=2\!\!\sum_{p_a+p_b>0}\!\!\frac{|\bra{a}\partial_\theta\rhoh\ket{b}|^2}{p_a+p_b},\qquad
\rhoh=\sum_a p_a\dyad{a},\label{eq:a_fq}
\end{equation}
which for a pure global state reduces to $\FQ=4(\braket{\Theta}{\Theta}-|\braket{\Psi_0}{\Theta}|^2)$. Keeping
the spin block, the boson block, or the joint block defines $\Fs,\Fb,\Fsb$ and the deficit
$\dFsb=\Fsb-\Fs-\Fb$ of Eq.~\eqref{eq:deficit_block}.

The global QFI is conserved. Since $\Hh$ is independent of $\theta$, both $\ket{\Psi_0}$ and
$\ket\Theta$ evolve under the same unitary, so every inner product between them is preserved and it is
enough to evaluate $\braket\Theta\Theta$ and $\braket{\Psi_0}{\Theta}$ at $t=0$. For the single-site
transverse encoding the latter is $\tfrac12\braket{\FM}{k}=0$; for the GHZ probe it is
$-i\langle\hat G\rangle$, which vanishes by symmetry. In both cases
$\Fglob=4\braket\Theta\Theta=4\,\mathrm{Var}_{\Psi_0}(\hat G)$ is set by the generator alone, the same for
every coupling and every time. Each block state is a partial trace of $\ket\Psi$, and
discarding part of a system cannot increase what is left to learn from it~\cite{Toth2014}, so no share
exceeds the conserved total. For one excitation the site-resolved shares are limited by something more
concrete still. Equation~\eqref{eq:a_fs} says each spin's share is its excitation weight multiplied by a
cloud overlap that never exceeds unity, so the site sum cannot exceed the total excitation weight, and the
same construction on a bath mode limits its share to the weight sitting in the bath. Those two weights are
the whole state for a single-site encoding, so the residual of Eq.~\eqref{eq:deficit} cannot come out negative
at the encoding point, and vanishes when the spin and bath marginals jointly exhaust the phase information without joint-access excess, as under
Jaynes--Cummings. That counting uses the single excitation and does not carry over to an encoding spread across
several sites.

The encoded state itself follows from the reference and the tangent. Since
$e^{-i\theta\Sy_k}\ket{\uparrow}_k=\cos\tfrac\theta2\ket{\uparrow}_k+\sin\tfrac\theta2\ket{\downarrow}_k$,
\begin{equation}
\ket{\Psi(\theta,t)}=\cos\tfrac\theta2\,\ket{\Psi_0}
+\sin\tfrac\theta2\Big(\sum_{j}\psi_j(t)\,\ket{j}\ket{\eta_j(t)}+\ket{\FM}\ket{\zeta(t)}\Big),
\label{eq:a_state}
\end{equation}
the stationary reference $\ket{\Psi_0}$ plus the excitation of Eq.~\eqref{eq:a_tan}. Tracing out the bath and
the spins outside site $j$ leaves the single-site density matrix in the basis
$\{\ket{\uparrow}_j,\ket{\downarrow}_j\}$; the boson sector $\ket{\FM}\ket{\zeta}$ raises the $\ket{\uparrow}_j$
population and adds $\sin^2\tfrac\theta2\,\psi_j\braket{\zeta}{\eta_j}$ to the coherence, which is
$O(\theta^2)$ and so drops out of the derivative at the encoding point (for the two couplings used here it
vanishes identically, $\ket{\eta_j}=\ket\varnothing$ with $\ket\zeta$ a one-boson state for
Jaynes--Cummings and $\ket\zeta=0$ for Holstein), so
\begin{equation}
\rhoh^{(1)}_j(\theta,t)=\begin{pmatrix} 1-p_j & C_j^{*}\\ C_j & p_j\end{pmatrix},\qquad
p_j=\sin^2\!\tfrac\theta2\,|\psi_j|^2,\qquad C_j=\tfrac12\sin\theta\,\psi_j\braket{\varnothing}{\eta_j}.
\label{eq:a_rho1}
\end{equation}
At $\theta=0$ this is $\mathrm{diag}(1,0)$ and, since $\braket{\Psi_0}{\Theta}=0$, the derivative
$\partial_\theta\rhoh^{(1)}_j|_0=\tfrac12\psi_j\braket{\varnothing}{\eta_j}\,|{\downarrow}\rangle\langle{\uparrow}|+\mathrm{h.c.}$
is purely off-diagonal. Equation~\eqref{eq:a_fq} then gives the exact block QFI density
\begin{equation}
\Fs(j,t)=|\psi_j(t)|^2\,\big|\braket{\varnothing}{\eta_j(t)}\big|^2,\label{eq:a_fs}
\end{equation}
the single-excitation amplitude weighted by the overlap of its accompanying bath state with the reference vacuum.
For Jaynes--Cummings the spin excitation and boson never coexist, $\ket{\eta_j}$ is the bath vacuum, and
$\braket{\varnothing}{\eta_j}=1$, so $\Fs(j,t)=|\psi_j|^2$ and the QFI deficit vanishes ($\dFsbbar=0$); the spin and bath marginals jointly exhaust the phase information, even though the state generically carries spin--bath entanglement.

The same construction with the whole spin register kept gives the full spin marginal. Its derivative at the
encoding point is again purely off-diagonal against the reference,
$\partial_\theta\rhoh^{\rm s}|_0=\tfrac12\sum_j\psi_j\braket{\varnothing}{\eta_j}\,\ket{j}\bra{\FM}+{\rm h.c.}$;
the boson sector contributes a diagonal term proportional to $\braket{\varnothing}{\zeta}$, which vanishes
for both couplings used here ($\ket\zeta$ a one-boson state for Jaynes--Cummings, $\ket\zeta=0$ for
Holstein). Equation~\eqref{eq:a_fq} then gives
$\FQ[\rhoh^{\rm s}]=\sum_j|\psi_j|^2|\braket{\varnothing}{\eta_j}|^2=\Fsbar$: for a single excitation the QFI
of the full spin marginal coincides with the site sum at the encoding point, so a collective spin measurement
has no more to read than site-resolved readout.

For Holstein the overlap is the source of the deficit, and it has a simple estimate. Approximating the
accompanying bath by a coherent cloud of $\bar n_j$ bosons gives the Franck--Condon factor
$|\braket{\varnothing}{\eta_j}|^2=e^{-\bar n_j}$. The occupation that enters here, and that is plotted in
Fig.~\ref{fig:hol}(b), is evaluated on the tangent state, since the reference $\ket\Psi$ carries no excitation at
$\theta=0$ and so drives no cloud; what the overlap needs is the bath the encoded branch actually pushes.
Summing Eq.~\eqref{eq:a_fs} with $\sum_j|\psi_j|^2=1$ and $\Fbbar=0$ gives
\begin{equation}\label{eq:fsbar_exact}
  \Fsbar=\sum_j|\psi_j|^2\,\big|\braket{\varnothing}{\eta_j}\big|^2
\end{equation}
exactly. Reaching Eq.~\eqref{eq:hol_deficit} from it takes two further steps, and both are approximations.
The first writes each overlap as $e^{-\bar n_j}$, which holds for a coherent cloud. The cloud accompanying a
moving excitation is not coherent: Eq.~\eqref{eq:a_eta} builds $\ket{\eta_j}$ from every path that brought the
excitation to site $j$, and those paths displace different modes. The second step gives every occupied
branch the same cloud size. Comparing the closed form with the simulation at the same instant, the deficit
runs a few percent above $1-e^{-\Nbtot}$, the excess growing as the bath disperses and reaching five percent (relative)
at $t_B/\omega_0=0.4$. That direction is the one a non-coherent cloud produces, since such a state overlaps
the vacuum less than a coherent state holding the same number of bosons. Equation~\eqref{eq:hol_deficit} is
accordingly a good estimate over the range computed here and not a limit the deficit must respect.

A product ansatz that gives the whole excitation superposition a single common bath captures the
occupations but not this entanglement; to reach the deficit it must be read branch by branch, each excitation
position carrying its own cloud.

\section{Jaynes--Cummings single-particle solution}\label{app:jc}
The Jaynes--Cummings
coupling conserves the total excitation number $\hat N_{\rm exc}=\sum_j(\nm_j+\ad_j\ah_j)$ and annihilates
the joint vacuum $\ket{\FM}\ket{\varnothing}$. A single excitation therefore lives at every instant either as
a spin excitation on one of the $L$ sites or as one boson in one of the $L$ modes, and the dynamics stays inside the
$2L$-dimensional single-particle sector. Writing the state in this basis,
\begin{equation}
\ket{\Psi(t)}=\sum_{j=1}^{L}a_j(t)\ket{j}\ket{\varnothing}+\sum_{j=1}^{L}b_j(t)\ket{\FM}\ket{1;j},\label{eq:b_state}
\end{equation}
the Schr\"odinger equation becomes a linear $2L\times2L$ problem $i\partial_t(a,b)^{\!\top}=H(a,b)^{\!\top}$,
\begin{equation}
H=\begin{pmatrix}H_{\rm m}&\lambda\Ident\\ \lambda\Ident&H_{\rm b}\end{pmatrix},\qquad
(H_{\rm m})_{j,j\pm1}=-\tfrac J2,\qquad (H_{\rm b})_{j,j\pm1}=-t_B,\qquad (H_{\rm b})_{jj}=\omega_0,\label{eq:b_H}
\end{equation}
whose diagonal blocks are the spin-excitation and boson tight-binding chains and whose off-diagonal blocks are the
on-site swap $\lambda$. Because the spin-excitation branch always carries the bare bath vacuum, the reduced single-site
state stays diagonal in the excitation position, and Eq.~\eqref{eq:a_fs} collapses to the excitation occupation,
\begin{equation}
\Fs(j,t)=|a_j(t)|^2,\qquad \Fsbar(t)=\sum_j|a_j|^2=1-\sum_j|b_j|^2.\label{eq:b_fs}
\end{equation}
The QFI is thus the probability that the excitation is still a spin excitation, its matter fraction, while the
complement has been swapped into the bath and is dark to a spin measurement.

\subsection{Reduction to two-level systems}
The swap acts on-site and conserves lattice momentum $q=2\pi m/L$, so a spin excitation of momentum $q$ couples only
to the boson of the same momentum. The $2L$ problem factorizes into $L$ independent two-level systems, one per
$q$, each pairing the bare spin-excitation energy $\varepsilon^m_q=-J\cos q$ with the bare boson energy
$\varepsilon^b_q=\omega_0-2t_B\cos q$. Diagonalizing the $2\times2$ block gives the two eigenenergies
\begin{equation}
E_\pm(q)=\bar\varepsilon_q\pm\tfrac12\Omega_q,\quad \bar\varepsilon_q=\tfrac12(\varepsilon^m_q+\varepsilon^b_q),\quad
\Omega_q=\sqrt{\delta_q^2+4\lambda^2},\quad \delta_q=\omega_0+(J-2t_B)\cos q,\label{eq:b_bands}
\end{equation}
split by the Rabi frequency $\Omega_q$ and mixed through an angle obeying
$|\tan2\vartheta_q|=2\lambda/|\delta_q|$. The
detuning $\delta_q$ sets the mixing. Far off resonance, $|\delta_q|\gg\lambda$, the branches are almost pure
spin excitation and pure boson and exchange little character; on resonance, $\delta_q=0$, they are equal
superpositions split by $2\lambda$ and the swap is complete. An initial spin excitation of momentum $q$ is a
superposition of the two branches, so it Rabi-oscillates between spin excitation and boson,
\begin{align}
a_q(t)&=e^{-i\bar\varepsilon_q t}\Big[\cos\tfrac{\Omega_q t}{2}+i\tfrac{\delta_q}{\Omega_q}\sin\tfrac{\Omega_q t}{2}\Big],
\qquad b_q(t)=-i\tfrac{2\lambda}{\Omega_q}e^{-i\bar\varepsilon_q t}\sin\tfrac{\Omega_q t}{2},\label{eq:b_aqbq}\\
|a_q(t)|^2&=1-\frac{4\lambda^2}{\Omega_q^2}\sin^2\tfrac{\Omega_q t}{2}.\label{eq:b_aq2}
\end{align}
A localized encoding excites every momentum with equal weight. Writing $a_q(t)$ for the single-channel
amplitude of Eq.~\eqref{eq:b_aqbq}, normalized to $a_q(0)=1$, the encoding at site $k$ distributes weight
$1/L$ over the $L$ channels, and since they evolve independently while the QFI counts only the
spin-excitation content the total is their average,
\begin{equation}
\Fsbar(t)=\frac1L\sum_q|a_q(t)|^2\ \xrightarrow{L\to\infty}\
1-4\lambda^2\!\int_{-\pi}^{\pi}\!\frac{dq}{2\pi}\,\frac{\sin^2(\Omega_q t/2)}{\Omega_q^2}.\label{eq:b_total}
\end{equation}
These momentum forms treat the chain as translationally invariant, whereas the model of
Eq.~\eqref{eq:H} is open. They are therefore the infinite-chain result, and they describe the finite open
chain only until the packet reaches an edge. That matters here because the encoding site sits at the edge already; for an edge encoding the reflected
amplitude returns immediately, and the open-chain $\Fsbar$ departs from
Eq.~\eqref{eq:b_total} by a few percent of the departed share within several inverse couplings and by roughly ten percent
thereafter, at every chain length checked. The parallel-band point $t_B=J/2$ is the exception: there the spin
and boson hopping matrices coincide, the swap factorizes from the propagation even with boundaries, and the
site-summed forms of Eqs.~\eqref{eq:b_parallel} and~\eqref{eq:b_res} hold exactly on the open chain, with the
Bessel profile of the spatial density replaced by the open-chain propagator. The closed forms below are used to identify mechanisms and limits; every
number quoted in the main text comes from the open chain.

The single-excitation QFI is a superposition of Rabi oscillations at the frequencies
$\Omega_q$, each of weight $4\lambda^2/\Omega_q^2$. Whether the share that has leaked into the bath
returns coherently or dephases depends on how the $\Omega_q$ are distributed across the band, and that
distribution is what $(\omega_0,t_B)$ control.

\subsection{Limiting regimes}
{\it (i) Short times.} At times short compared with every $1/\Omega_q$, the integrand expands as
$\sin^2(\Omega_q t/2)/\Omega_q^2=t^2/4-\Omega_q^2t^4/48+O(t^6)$, and averaging with
$\langle\Omega_q^2\rangle=4\lambda^2+\omega_0^2+\tfrac12(J-2t_B)^2$ gives
\begin{equation}
\Fsbar(t)=1-\lambda^2t^2+\frac{\lambda^2}{12}\Big(4\lambda^2+\omega_0^2+\tfrac12(J-2t_B)^2\Big)t^4+O(t^6).\label{eq:b_short}
\end{equation}
The leading term $-\lambda^2t^2$ is set by the coupling alone, and the bath enters only at fourth order
through $\langle\Omega_q^2\rangle$. The initial decay coefficient of the QFI is therefore bath-independent,
fixed by the bare swap rate.

{\it (ii) Parallel bands ($t_B=J/2$).} Here $\delta_q=\omega_0$ is momentum-independent, so a single frequency
$\Omega=\sqrt{\omega_0^2+4\lambda^2}$ governs every channel and, with $\sum_j J_{j-k}^2(Jt)=1$, the space--time
dependence factorizes,
\begin{equation}
\Fs(j,t)=\Big[1-\tfrac{4\lambda^2}{\Omega^2}\sin^2\tfrac{\Omega t}{2}\Big]J_{j-k}^2(Jt),\qquad
\Fsbar(t)=1-\frac{4\lambda^2}{\omega_0^2+4\lambda^2}\sin^2\tfrac{\Omega t}{2},\qquad T=\frac{2\pi}{\Omega}.\label{eq:b_parallel}
\end{equation}

{\it (iii) Resonance ($\omega_0=0,\,t_B=J/2$).} Then $\delta_q=0$ and $\Omega=2\lambda$, so the swap is complete,
\begin{equation}
\Fsbar(t)=\cos^2(\lambda t),\qquad T=\pi/\lambda.\label{eq:b_res}
\end{equation}

{\it (iv) Antiadiabatic ($\omega_0\gg J,\lambda$).} Every channel is far off resonance, $\Omega_q\to\omega_0$
and the swap weight $4\lambda^2/\omega_0^2\to0$,
\begin{equation}
\Fsbar(t)\to1-\frac{4\lambda^2}{\omega_0^2}\sin^2\tfrac{\omega_0 t}{2}.\label{eq:b_anti}
\end{equation}

{\it (v) Detuned ($t_B\ne J/2$).} $\Omega_q$ varies across the zone and the channels dephase; the time average
$\sin^2\to\tfrac12$ gives the long-time plateau
\begin{equation}
\bar{\mathcal F}^{\rm s}_{\infty}=1-2\lambda^2\!\int_{-\pi}^{\pi}\!\frac{dq}{2\pi}\,\frac{1}{\Omega_q^2}.\label{eq:b_plateau}
\end{equation}
The physics of these limits is discussed in Sec.~\ref{sec:jc}.

\subsection{Perfect-transfer Hamiltonian}
The engineered profile breaks translation invariance, so momentum no longer labels the modes of the chain of
Sec.~\ref{sec:pst}. The profile $J_i\propto\sqrt{i(L-i)}$ makes the single-excitation block of the Hamiltonian the $\hat J_x$ operator of a
spin $j=(L-1)/2$, $H_{\rm m}=-s\hat J_x$ with $s=2J/L$, so the excitation spectrum is the equally spaced ladder
$\varepsilon^m_m=-s\,m$, $m=-j,\dots,j$, and the eigenvectors are the real Wigner rotation matrix
$V_{nm}=d^{\,j}_{nm}(\pi/2)$. Evolution for a time $t$ is a rotation by $st$ about $\hat J_x$, and at
$t_\star=\pi/s=\pi L/2J$ it inverts $\hat J_z$ and mirrors the chain~\cite{Christandl2004}, in every
excitation sector at once~\cite{Albanese2004}.

At $t_B=0$ every bath mode has the same energy, so the ladder diagonalizes the boson block as well and the
on-site swap couples each rung only to the boson standing on the same rung. A spin excitation on rung $m$ therefore
Rabi-oscillates into that boson and back, detuned by $\delta_m=\omega_0+s\,m$ and split by
$\Omega_m$, exactly as a spin excitation of momentum $q$ does on the uniform chain, and the mirror carries whatever
amplitude has stayed a spin excitation. Hence
\begin{equation}
f_{n,k}(t)=\sum_{m=-j}^{j}V_{nm}\,a_m(t)\,V_{km},\qquad
a_m(t)=e^{-i\bar\varepsilon_m t}\Big[\cos\tfrac{\Omega_m t}{2}+i\tfrac{\delta_m}{\Omega_m}\sin\tfrac{\Omega_m t}{2}\Big],
\label{eq:b_pst}
\end{equation}
exact at every $\lambda$ and $\omega_0$, agreeing with a direct exponentiation of Eq.~\eqref{eq:b_H} to
$10^{-14}$ and with the many-body simulation of Sec.~\ref{sec:pst} to $2\times10^{-4}$.

At $\lambda=0$ each $a_m$ reduces to a bare phase and Eq.~\eqref{eq:b_pst} returns the mirror,
$|f_{L,1}(t_\star)|^2=1$. When $\lambda$ dominates the bandwidth the detunings are negligible beside it and
$\Omega_m\to2\lambda$ on every rung, so the rung sum factorizes into
$f_{n,k}(t)\to e^{-i\omega_0t/2}\cos(\lambda t)\,f^{(0)}_{n,k}(t/2)$, a Rabi envelope on a mirror running at
half rate, which is the static-bath result of Ref.~\cite{Burgarth2006} and places the arrival at $2t_\star$.
At $\lambda/\omega_0=0.5$ the coupling reaches $0.29$ of the bandwidth and the arrival sits at
$1.18\,t_\star$, moving to $1.99\,t_\star$ only once $\lambda$ passes the band by an order of magnitude.
The engineered chain interpolates between perfect transfer and the half-velocity limit, and
Sec.~\ref{sec:pst} works between them.

The Holstein coupling commutes with the excitation number and not with the boson number, so where the swap
relocates a single excitation between a spin and a mode, the displacement $\nm_i(\ad_i+\ah_i)$ raises bosons
out of the vacuum at every hop and raises them in unbounded number. The state carries weight in every boson
sector at once, and the counting that produced Eq.~\eqref{eq:b_H}, one excitation lying either on a spin or in
a mode, is left with nothing to count. An excitation carrying a cloud of indefinite size through a lattice is the
self-trapping problem, and the many-body simulation of Sec.~\ref{sec:pst} is what reaches it. The
Jaynes--Cummings coupling solved here instead conserves the total excitation, which both lets the phase into the
bath and keeps the single-excitation sector finite.

\section{Register fidelity and block QFI of the arriving probe}\label{app:fidelity}
Both quantities of Eq.~\eqref{eq:fid_vs_qfi} are fixed by the three numbers of Eq.~\eqref{eq:ghz_sector}. Write
$\ket{{\rm d}}=\ket{\downarrow^n}$, $\ket{{\rm u}}=\ket{\uparrow^n}$ and
$\rhoh^{{\rm s},(n)}_{\rm du}=c\,e^{i\varphi}$. The target register is
$\ket{{\rm GHZ}(\chi)}=\tfrac1{\sqrt2}(\ket{{\rm d}}+e^{i\chi}\ket{{\rm u}})$, so
\begin{equation}
  F(\chi)=\bra{{\rm GHZ}(\chi)}\rhoh^{{\rm s},(n)}\ket{{\rm GHZ}(\chi)}
  =\tfrac12\big(p_{\rm d}+p_{\rm u}\big)+c\cos(\chi+\varphi),
\end{equation}
which at the phase the transfer actually delivers, $\chi=-\varphi$, gives the first of
Eqs.~\eqref{eq:fid_vs_qfi}. Only the cross term carries $c$, and it enters once.

For the QFI, $\hat G$ acts on the two states as $\hat G\ket{{\rm u}}=\tfrac n2\ket{{\rm u}}$ and
$\hat G\ket{{\rm d}}=-\tfrac n2\ket{{\rm d}}$, so encoding sends $\rhoh_{\rm du}\to\rhoh_{\rm du}e^{-in\theta}$
and leaves every population unchanged. The derivative is therefore purely off-diagonal in this sector,
$|\partial_\theta\rhoh_{\rm du}|=n\,c$, and the spectral form~\eqref{eq:a_fq} keeps only that one pair,
\begin{equation}
  \Fs(1,t;n)=2\,\frac{|\partial_\theta\rhoh_{\rm du}|^2}{p_{\rm d}+p_{\rm u}}
            +2\,\frac{|\partial_\theta\rhoh_{\rm ud}|^2}{p_{\rm u}+p_{\rm d}}
            =\frac{4n^2c^2}{p_{\rm u}+p_{\rm d}},
\end{equation}
the second of Eqs.~\eqref{eq:fid_vs_qfi}. We justify this reduction in two steps.

First, restricting to the pair $\ket{{\rm d}},\ket{{\rm u}}$ is not a matter of the other configurations
lacking coherence with them. The reduction is exact because of a superselection rule. The coupling conserves
the total number of spin excitations, and the two launched branches carry the definite numbers $0$ and $n$,
so a block coherence between configurations with $k$ and $k'$ excitations survives the trace over the rest
of the chain only for $k=k'$, the single $\ket{{\rm d}}\!-\!\ket{{\rm u}}$ element excepted. Both the $0$- and the
$n$-excitation sectors of an $n$-spin block are one-dimensional, so that pair is exactly the sector carrying
$\theta$. Without number conservation the reduction fails.

Second, $\ket{{\rm d}}$ and $\ket{{\rm u}}$ are eigenvectors of $\hat G$ but not of $\rhoh^{{\rm s},(n)}$,
while Eq.~\eqref{eq:a_fq} is written in the eigenbasis of the state. It gives the same answer here because
$\partial_\theta\rhoh$ is proportional to $\sigma_y$ in the $({\rm d},{\rm u})$ basis whereas the Bloch
vector of the block lies in the $x$--$z$ plane; the derivative therefore stays purely off-diagonal in the true
eigenbasis with the same modulus $nc$, and the denominator is the trace of the pair, which no change of basis
alters. The denominator is that trace and not unity, because the block also holds partially transferred
configurations, which carry population but no $\theta$ dependence. Away from the encoding point the
populations stay $\theta$-independent, so both forms hold for the whole family.

The QFI scales as $c^2$ while the fidelity scales as $c$, so the two diverge whenever the channel removes coherence faster than it removes weight. Writing
$\bar p=\tfrac12(p_{\rm u}+p_{\rm d})$ for the weight the branches retain, the ratio of the readable share to
the fidelity is $2c^2/[\bar p(\bar p+c)]$. A state cannot carry more coherence than weight, $c$ never
exceeding $\bar p$, so this ratio never rises above one and the readable share sits at or below the fidelity
however the channel behaves. The two coincide when the retained weight is a clean pair of GHZ branches,
$c=\bar p$, and both reach unity only if nothing is lost at all. The two therefore separate according to
whether the channel removes coherence faster than weight, as a dephasing bath does. The many-body simulation supplies
$p_{\rm u}$, $p_{\rm d}$ and $c$; the two expressions above are exact within the two-dimensional sector
spanned by $\ket{{\rm d}}$ and $\ket{{\rm u}}$. The block carries weight outside that sector as well,
$p_{\rm d}+p_{\rm u}$ falling from $0.91$ to $0.71$ between $L=8$ and $L=20$ here, so against the QFI
extracted from the whole $2^{n}$-dimensional block they are an approximation whose residual grows with the
chain, from $5\times10^{-7}$ at $L=8$ to $2\times10^{-5}$ at $L=20$.

\section{Exact solution of the frozen GHZ probe}\label{app:ghz_exact}
When the Holstein probe occupies every site, $L=n$, the model is solved outright and no ansatz enters. The exchange annihilates both branches of the
GHZ state, $\Hh_{\rm s}\ket{\uparrow}^{\otimes n}=\Hh_{\rm s}\ket{\downarrow}^{\otimes n}=0$, and the Holstein
coupling conserves the excitation number, so each branch remains the polarization eigenstate it started as and
the excitation configuration never changes. The dynamics therefore separates,
\begin{equation}
  \ket{\Psi(t)}=\tfrac{1}{\sqrt2}\Big(e^{i\varphi(t)}\ket{\downarrow}^{\otimes n}\ket{\bm\beta(t)}
  +\ket{\uparrow}^{\otimes n}\ket{\varnothing}\Big),
  \label{eq:ghz_state_frozen}
\end{equation}
and each branch is evolved by its own bath Hamiltonian at fixed $\nm_l$. In the polarized branch
$\nm_l=0$ and the bath is untouched. In the excited branch $\nm_l=1$ on every site, so that bath
Hamiltonian is
$\sum_{ll'}h_{ll'}\ad_l\ah_{l'}+\lambda\sum_l(\ah_l+\ad_l)$, quadratic with a constant linear drive. Such a
Hamiltonian displaces the vacuum into a coherent state and nothing else, which is why $\ket{\bm\beta(t)}$ is
the exact state of that branch, explaining why $\varphi(t)$ is a c-number that drops
out below. Writing
$h_{ll'}=\omega_0\delta_{ll'}-t_B(\delta_{l,l'+1}+\delta_{l,l'-1})$ for the bath matrix and $\bm 1$ for the
vector of ones,
\begin{equation}
  \bm\beta(t)=\lambda\,h^{-1}\big(e^{-iht}-\mathbb{I}\big)\bm 1,\qquad
  \Nbcond(t)=|\bm\beta(t)|^2,
  \label{eq:ghz_exact}
\end{equation}
which solves $i\dot{\bm\beta}=h\bm\beta+\lambda\bm 1$, the Heisenberg equation of that branch Hamiltonian
for $\langle\ah_l\rangle$, from $\bm\beta(0)=0$. Here $\Nbcond$ is the occupation of the bath state
conditional on the displaced branch; only one of the two branches is displaced, so the occupation of the
state as a whole is $\Nbtot=\tfrac12\Nbcond$, and it is the conditional occupation that sets the overlap
below.
Tracing the bath out of Eq.~\eqref{eq:ghz_state_frozen} leaves the two branches with weight $\tfrac12$ each and a
single off-diagonal element, so the spin block is the two-level state
\begin{equation}
  \rhoh^{\rm s}=\tfrac12\Big(\dyad{\uparrow^{n}}+\dyad{\downarrow^{n}}
  +e^{i\varphi}\braket{\varnothing}{\bm\beta}\,\ket{\uparrow^{n}}\bra{\downarrow^{n}}+{\rm h.c.}\Big),
  \label{eq:ghz_rhos}
\end{equation}
whose coherence is the overlap of the two conditional bath states,
$\braket{\varnothing}{\bm\beta}=e^{-|\bm\beta|^2/2}$ for a coherent state. Both branches are eigenstates of
the generator, $\hat G\ket{\uparrow}^{\otimes n}=\tfrac n2\ket{\uparrow}^{\otimes n}$ and
$\hat G\ket{\downarrow}^{\otimes n}=-\tfrac n2\ket{\downarrow}^{\otimes n}$, so encoding sends
$c\to c\,e^{-i\theta n}$ with $c=\tfrac12 e^{i\varphi}\braket{\varnothing}{\bm\beta}$ and leaves the
populations alone. Equation~\eqref{eq:ghz_rhos} is then a qubit whose Bloch vector has length $2|c|$ and is
rotated at rate $n$ in the equatorial plane, so $\partial_\theta\bm r\perp\bm r$, the second term of the
qubit QFI drops, and $\FQ=|\partial_\theta\bm r|^2=4n^2|c|^2$. Hence
\begin{equation}
  \Fs(1,t;n)=n^2\,e^{-|\bm\beta(t)|^2}.
  \label{eq:ghz_exact_fs}
\end{equation}
For an Einstein bath this is explicit. The modes decouple at $t_B=0$, $h=\omega_0\Ident$, and every occupied
site is driven identically, $\beta_l=(\lambda/\omega_0)(e^{-i\omega_0t}-1)$, so
$|\bm\beta|^2=2ng(1-\cos\omega_0t)$ with $g=(\lambda/\omega_0)^2$ and
\begin{equation}
  \Fs(1,t;n)/n^2=\exp\!\big[-2ng\,(1-\cos\omega_0 t)\big]=\big[e^{-2g(1-\cos\omega_0 t)}\big]^{n},
  \label{eq:ghz_einstein}
\end{equation}
which is Eq.~\eqref{eq:ghz_hol_frozen}. The exponent vanishes at $t=2\pi k/\omega_0$ for every $n$, so the
revival is exact and independent of the probe size, while the deepest suppression, $e^{-4ng}$, falls
exponentially with it. Equation~\eqref{eq:ghz_einstein} is the register decoherence function of
Ref.~\cite{Reina2002} in its independent-decoherence limit; the QFI interpretation and the three-party
sensitivity partition are new here.
The branches remain polarization eigenstates for as long as the coupling conserves the excitation number, so the
exchange annihilates them whatever its strength, making Eq.~\eqref{eq:ghz_einstein} valid at any $J$. It differs
from the Jaynes--Cummings result of Eq.~\eqref{eq:ghz_frozen}, which the exchange spoils as soon as a spin is
raised in the lower branch.
The Franck--Condon form of Eq.~\eqref{eq:hol_deficit} holds exactly here, since the frozen probe reduces to
the independent-boson model~\cite{Duke1965,Mahan2000,Palma1996}, with the conditional occupation $\Nbcond$
as the cloud occupancy in the exponent. In the single-excitation sector of Sec.~\ref{sec:hol} the excited branch carries
the whole weight, so $\Nbcond=\Nbtot$; for the GHZ probe the two differ by the branch weight.
Section~\ref{sec:hol} shows that the same form survives when the excitation is free to move and carry its cloud,
which Eq.~\eqref{eq:ghz_exact_fs} cannot address. Equation~\eqref{eq:ghz_exact} also fixes the boson cutoff
analytically, a task the conserved $\Fglob$ cannot perform. The bath spectrum
$\omega_k=\omega_0-2t_B\cos[k\pi/(L+1)]$ softens as $t_B\to\omega_0/2$, the displacement $|\bm\beta|$ grows
like $1/\omega_{\rm min}$, and at $\lambda=0.6$, $\omega_0=1$, $n=3$ the peak of the conditional occupation
$\Nbcond$, the occupancy the cutoff must accommodate, climbs from $4.3$ at $t_B=0$ to $12.7$ at $t_B=0.3$
and $49$ toward the soft-mode limit at $t_B=0.5$, the discrete
$n$-mode spectrum staying bounded below throughout.

\section{Probe-angle dependence of the bath Fisher information}\label{app:theta}
\begin{figure}[h]
  \centering
  \includegraphics[width=\columnwidth]{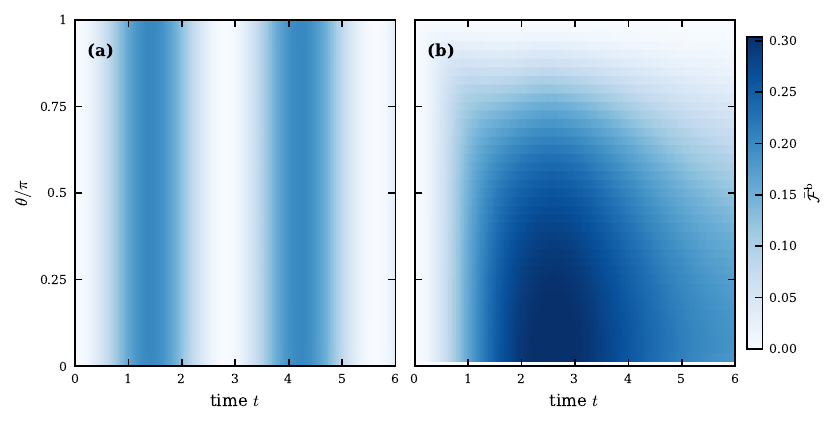}
  \caption{Probe-angle dependence of the site-summed bath QFI $\bar{\mathcal{F}}^{\mathrm{b}}(\theta, t)$
  for (a)~Jaynes--Cummings and (b)~Holstein coupling [$L=20$, $\omega_0=2J$, $t_B=0.5J$, $\lambda=0.5J$,
  $N_{\max}=3$; TEBD, $dt=0.05$, $41$ angles across $[0,\pi]$]. The Jaynes--Cummings response is independent
  of $\theta$. The Holstein bath QFI vanishes exactly along $\theta=0$ and $\theta=\pi$ for all $t$. At
  $\theta=0$ the QFI is discontinuous~\cite{Safranek2017,Seveso2020}, reaching the order of the correlation
  deficit already at the first angle of the scan, and it decreases monotonically to zero as
  $\theta\to\pi$.}
  \label{fig:theta_heatmap}
\end{figure}

Figure~\ref{fig:theta_heatmap} shows $\bar{\mathcal{F}}^{\mathrm{b}}=\sum_j\Fb(j,t)$, the site-summed bath QFI,
as a function of time and probe angle $\theta$ for the Jaynes--Cummings and Holstein couplings, at
$L=20$, $\omega_0=2J$, $t_B=0.5J$, $\lambda=0.5J$, $N_{\max}=3$.
For Jaynes--Cummings [panel~(a)] the bath QFI is independent of $\theta$, and the independence is exact.
Tracing all spins and every mode but $l$ from the encoded state [Eq.~\eqref{eq:a_state}]
leaves a two-level marginal in the number basis $\{\ket{0;l},\ket{1;l}\}$,
\begin{equation}
\rhoh^{{\rm b},(1)}_l(\theta,t)=\begin{pmatrix}1-p_l & C_l^{*}\\ C_l & p_l\end{pmatrix},\qquad
p_l=\sin^2\!\tfrac\theta2\,|b_l|^2,\qquad C_l=\tfrac12\sin\theta\,b_l,
\label{eq:e_rhol}
\end{equation}
with $b_l$ the boson amplitude of Eq.~\eqref{eq:jc_state}; the spin-excitation components and the bosons on
other modes carry orthogonal environments and feed only the $\ket{0;l}$ population. The phase of $b_l$ does
not depend on $\theta$, so the Bloch vector of the family moves in a fixed plane through the $z$ axis, with
in-plane components
\begin{equation}
\bm r_l(\theta)=\Big(|b_l|\sin\theta,\;1-|b_l|^2\,(1-\cos\theta)\Big),\qquad
1-|\bm r_l|^2=w(1-w)(1-\cos\theta)^2,\qquad w\equiv|b_l|^2,
\label{eq:e_bloch}
\end{equation}
and the qubit QFI $\FQ=|\partial_\theta\bm r|^2+(\bm r\cdot\partial_\theta\bm r)^2/(1-|\bm r|^2)$ evaluates to
\begin{equation}
\FQ\big[\rhoh^{{\rm b},(1)}_l\big]
=\underbrace{w\cos^2\theta+w^2\sin^2\theta}_{|\partial_\theta\bm r|^2}
+\underbrace{w(1-w)\sin^2\theta}_{\text{mixed-state term}}
= w ,
\label{eq:e_qfi}
\end{equation}
using $\bm r\cdot\partial_\theta\bm r=-w(1-w)(1-\cos\theta)\sin\theta$; the pure boundary cases
($\theta=0,\pi$ or $w\in\{0,1\}$) follow by continuity. The $\theta$-dependence of the Bloch-vector speed
cancels against the mixed-state correction, leaving
\begin{equation}
\Fb(l,t)=|b_l(t)|^2\quad\text{for every }\theta,\qquad
\bar{\mathcal{F}}^{\rm b}(\theta,t)=\sum_l|b_l(t)|^2,
\label{eq:e_sum}
\end{equation}
the swapped weight of Eq.~\eqref{eq:jc_main}, for every probe direction; the temporal oscillations in
panel~(a) reflect the Rabi-like exchange between the spin and bath sectors.
For Holstein [panel~(b)] the bath QFI vanishes identically along $\theta=0$ and $\theta=\pi$ for all $t$.
The Holstein coupling [Eq.~\eqref{eq:hol}] commutes with the spin-excitation number $\hat{N}^{\mathrm{m}}$,
so the two branches of the encoded state,
\begin{equation}
\ket{\Psi(\theta,t)}=\cos\tfrac\theta2\,\ket{\FM}\ket{\varnothing}+\sin\tfrac\theta2\,\ket{\Phi(t)},\qquad
\ket{\Phi(t)}=e^{-i\Hh t}\ket{k}\ket{\varnothing},
\label{eq:e_branches}
\end{equation}
evolve independently and carry the definite spin-excitation numbers $0$ and $1$. Tracing out the spins kills
every cross term between them, so the bath marginal is a classical mixture whose entire $\theta$-dependence
sits in the weights,
\begin{equation}
\rhoh^{\rm b}(\theta,t)=\cos^2\!\tfrac\theta2\,\dyad{\varnothing}+\sin^2\!\tfrac\theta2\,\rhoh^{\rm b}_\Phi(t),
\qquad \rhoh^{\rm b}_\Phi=\Tr_{\rm spin}\dyad{\Phi},
\label{eq:e_mix}
\end{equation}
and its derivative
\begin{equation}
\partial_\theta\rhoh^{\rm b}(\theta,t)=\tfrac12\sin\theta\,\Big[\rhoh^{\rm b}_\Phi(t)-\dyad{\varnothing}\Big]
\label{eq:e_deriv}
\end{equation}
vanishes on both lines, $\theta=0$ and $\theta=\pi$. The zeros are therefore exact and hold for any
$\lambda$, $t_B$, or $L$; the argument applies unchanged to every mode marginal, so each term of the site
sum vanishes with it. The approach to the two zeros is sharply asymmetric. At $\theta=0$ the mixture of
Eq.~\eqref{eq:e_mix} changes rank, and the QFI of a rank-changing family is discontinuous at the rank-change
point~\cite{Safranek2017,Seveso2020}. Already at the smallest nonzero angle of the scan
($\theta=\pi/40$) the site-summed bath QFI reaches $\simeq0.31$, comparable to the correlation deficit
[cf.\ Eq.~\eqref{eq:hol_deficit}]; it is nearly constant at small $\theta$ and decreases monotonically
toward $\theta=\pi$, where it vanishes continuously, since the vacuum branch that a bath measurement
would have to single out lies almost entirely inside the support of the cloud-bearing branch. The scan uses $N_{\max}=3$; increasing it to $N_{\max}=5$ leaves $\bar{\mathcal{F}}^{\rm b}(\theta,t)$ unchanged to plotting accuracy. The symmetry therefore protects only the first-order term at $\theta=0$; at any finite $\theta$ the
bath modes carry information about $|\theta|$ but none about its sign.

\section{Numerical certification of the tensor-network results}\label{app:convergence}
\begin{figure}[!ht]
  \centering
  \includegraphics[width=0.86\columnwidth]{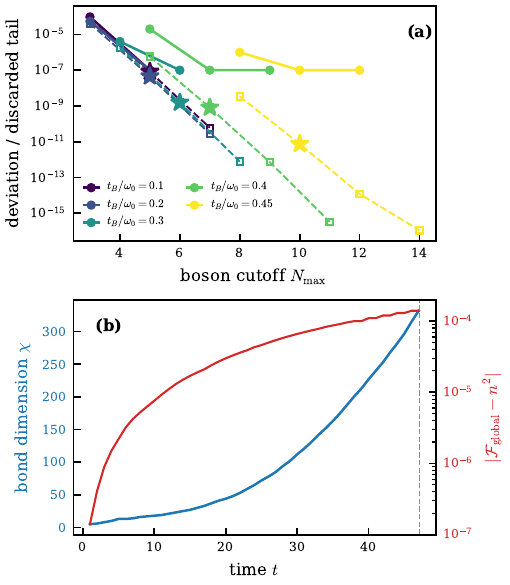}
  \caption{Numerical certification of the tensor-network results. (a)~Boson-cutoff convergence of the
  soft-mode sweep [Fig.~\ref{fig:softmode}; $L=20$, $\omega_0=2J$, $\lambda=0.5J$; TEBD]: deviation of the
  peak whole-chain deficit from its value at the largest cutoff computed (circles) and the discarded
  Poisson tail of the driven-oscillator estimate (open squares) vs the cutoff $N_{\max}$, for
  $t_B/\omega_0=0.1$--$0.45$; stars mark the production cutoffs, and deviations are floored at the printed
  precision $10^{-7}$. (b)~The $L=30$ Holstein transport run [$n=3$, $\omega_0=J$, $\lambda/\omega_0=0.1$,
  $t_B/\omega_0=0.3$, $N_{\max}=4$, $\chi\le1024$; TEBD, $dt=0.1$]: bond dimension $\chi(t)$ (left axis)
  and the drift $|\Fglob-n^2|$ of the conserved global QFI (right axis, log scale) up to the mirror time
  $t_\star=15\pi$ (dashed).}
  \label{fig:convergence}
\end{figure}

Two of the tensor-network computations push the method hardest: the soft-mode sweep of
Fig.~\ref{fig:softmode}, where the driven occupation grows as the band softens, and the engineered-chain
transport of Sec.~\ref{sec:pst}, where the entangled block crosses up to thirty sites.
Figure~\ref{fig:convergence} certifies both.

Panel~(a) addresses the boson cutoff on the hardest observable of the paper, the peak whole-chain deficit
of the soft-mode sweep. For each bath hopping, the circles show the deviation of the peak deficit from its
value at the largest cutoff computed, and the open squares the discarded Poisson tail of the
driven-oscillator estimate, against $N_{\max}$; stars mark the cutoff used in production. The deviation is
below $10^{-4}$ already at the smallest cutoffs shown, falls with $N_{\max}$ along the predicted tail, and
lies below the $10^{-7}$ printed precision at every production cutoff, the required cutoff growing as the
band softens, in line with the analytic occupation of App.~\ref{app:ghz_exact}.

Panel~(b) documents the longest transport run quoted in Sec.~\ref{sec:pst}, the $L=30$ Holstein GHZ
transport at the parameters of Fig.~\ref{fig:pst_holstein_L20}. The bond dimension grows to $\chi=332$ at
the mirror time $t_\star=15\pi$ without reaching the cap, and the drift of the analytically conserved
global QFI stays below $1.4\times10^{-4}$ out of $\Fglob=n^2=9$, certifying the bond truncation over the
whole crossing; the peak site occupation, $0.10$ boson, sits far inside the cutoff. The conserved total is
the natural certificate here, since it is exactly time-independent [App.~\ref{app:method}] and any
truncation error accumulates in it.

\end{document}